\documentclass[aps,pra,twocolumn,a4paper,
showpacs,preprintnumbers] {revtex4-1}
\usepackage{graphicx, color}
\usepackage[dvipsnames]{xcolor}
\usepackage{calc}
\usepackage{amsmath}
\usepackage{amssymb}
\usepackage{amsbsy}
\usepackage{enumitem}

\newcommand{\D}{{\rm{d}}}
\newcommand{\I}{{\rm{I}}}

\begin{document}
\preprint{PHYSICAL REVIEW A \textbf{96}, 043856 (2017)}

%\title{Free-space quantum links with haze, rain, and turbulence }
\title{Free-space quantum links under diverse weather conditions}

 \author{D. Vasylyev}
 \affiliation{Institut f\"ur Physik, Universit\"at Rostock,
 Albert-Einstein-Stra\ss e 23, 18059 Rostock, Germany}

 \author{ A. A. Semenov}
\affiliation{Institut f\"ur Physik, Universit\"at Rostock,
 Albert-Einstein-Stra\ss e 23, 18059 Rostock, Germany}

 \author{W. Vogel}
 \affiliation{Institut f\"ur Physik, Universit\"at Rostock,
  Albert-Einstein-Stra\ss e 23, 18059 Rostock, Germany}

   \author{K.~G\"unthner}
 \affiliation{Max Planck Institute for the Science of Light, Staudtstra\ss{}e 2, 91058 Erlangen, Germany}
\affiliation{Institut f\"ur Optik, Information und Photonik, Universit\"at Erlangen-N\"urnberg,
Staudtstra\ss{}e  7/B2, 91058 Erlangen, Germany}

	\author{A.~Thurn}
 \affiliation{Max Planck Institute for the Science of Light, Staudtstra\ss{}e 2, 91058 Erlangen,
 Germany}
 \affiliation{Institut f\"ur Optik, Information und Photonik, Universit\"at Erlangen-N\"urnberg,
 Staudtstra\ss{}e  7/B2, 91058 Erlangen, Germany}

 \author{\"O.~Bayraktar}
 \affiliation{Max Planck Institute for the Science of Light, Staudtstra\ss{}e 2, 91058 Erlangen,
 Germany}
 \affiliation{Institut f\"ur Optik, Information und Photonik, Universit\"at Erlangen-N\"urnberg,
 Staudtstra\ss{}e  7/B2, 91058 Erlangen, Germany}

     \author{Ch.~Marquardt}
 \affiliation{Max Planck Institute for the Science of Light, Staudtstra\ss{}e 2, 91058 Erlangen,
 Germany}
 \affiliation{Institut f\"ur Optik, Information und Photonik, Universit\"at Erlangen-N\"urnberg,
 Staudtstra\ss{}e  7/B2, 91058 Erlangen, Germany}

\begin{abstract}
Free-space optical communication links are  promising channels for establishing secure quantum communication. Here we study the transmission of nonclassical light through a turbulent atmospheric link under diverse weather conditions, including rain or haze.  To include these effects, the theory of light transmission through atmospheric links in the elliptic-beam approximation
presented by Vasylyev \textit{et al.} [D. Vasylyev \textit{et al.}, Phys. Rev. Lett. \textbf{117}, 090501 (2016); arXiv:1604.01373]
is further generalized.
It is demonstrated, with good agreement between theory and experiment, that low-intensity rain merely contributes additional deterministic losses, whereas haze also introduces additional  beam deformations of the transmitted light. Based on these results, we study theoretically  the transmission of quadrature squeezing and Gaussian entanglement under these weather conditions.
\end{abstract}

\maketitle

\section{Introduction}

By the use of modern quantum communication technologies, secret information exchange becomes  feasible~\cite{Skarani}. Light is the most attractive candidate for the practical use in quantum communication protocols due  to its robustness against the influence of the environment, high bandwidth, and the possibilities of multiplexing information encoding using, e.g., polarization or orbital angular momentum. Recently, quantum communication technologies in free-space channels have developed rapidly. Various experiments have been performed, with distances ranging from intracity~\cite{Resch, Martinez, Peuntinger, Krenn, Croal, Endo} to more than 100 km~\cite{Manderbach, Fedrizzi2009, Yin, Capraro, Herbst}, which have shown that the atmosphere is a reliable medium for the communication with  light exploiting its quantum properties. Moreover, the use of satellite-mediated links~\cite{Wang, Bourgoin, Vallone2015, Vallone2016, Guenthner, Liao, GangRen, Takenaka} paves the way to establish global
quantum communication links.

During the propagation through the atmosphere, optical beams undergo random  broadening and deformation as well as stochastic deflections as a whole.
 The major effect comes from the  turbulent fluctuations  of the refractive index. Moreover, the light beam may be attenuated by backward scattering and absorption.  These effects become even more pronounced if the meteorological conditions deviate from what we usually call clear weather. Indeed, under bad weather conditions the optical beam experiences additional broadening, absorption, and backscattering due to random scattering on dust particles, aerosols, and/or precipitations. There exist numerous studies of classical light propagation in  turbulent atmosphere in the presence of haze, fog, or rain~\cite{Hong, Liu, Deepak, Grabner,Lukin81, Yura, Lukin}. Some recent advances were made in the quantum theory of light
propagation through
atmospheric turbulence~\cite{Diament, Perina, Perina1973, Milonni, Paterson, Berman, Semenov, Vasylyev2012, Vasylyev2016, Chumak, Bohmann17}. However, a consistent quantum theory of light propagation in turbulence with the inclusion of random scattering on particles, aerosols,  and precipitations has not been developed so far.  In contrast to the theory of random scattering of  classical light \cite{Ishimaru, Hulst, IshimaruBook}, the quantum theory implies  additional constraints, such as commutation rules, so that the classical results cannot be applied in a straightforward manner.

The purpose of our  article is  to develop a basic quantum theory of
 light propagation in haze, rain,  and turbulence in close relation to experimental observations.  To this end, we relate the transmitted quantum state to the initial state at the transmitter site via the input-output relation
\begin{align}
 \hat a_{\mathrm{out}}=\sqrt{\eta}\hat a_{\mathrm{in}}+\sqrt{1-\eta}\hat c, \label{inout}
\end{align}
where $\hat a_{\mathrm{in(out)}}$ is the input(output) field annihilation operator and $\hat c$ is the operator of environmental modes.  The transmittance $\eta{\in}[0,1]$ is a random variable  that describes the fluctuating-loss channels under study. In terms   of the Glauber-Sudarshan $P$ function \cite{Glauber, Sudarshan}, the input-output relation (\ref{inout}) reads
\begin{align}\label{Pinout}
 P_{\rm out}(\alpha)=\int_0^1\D\eta\mathcal{P}(\eta)\frac{1}{\eta}P_{\rm in}\left(\frac{\alpha}{\sqrt{\eta}}\right).
\end{align}
Here the  functions $P_{\rm in}(\alpha)$ and $P_{\rm out}(\alpha)$ are quasiprobability distributions that   completely describe the input and output quantum states, respectively. The probability distribution of the transmittance (PDT) $\mathcal{P}(\eta)$
describes fluctuations of the transmission efficiency $\eta$.

In many practical situations one deals with phase-insensitive measurements. Hence, the phase of $\eta$ is not needed in the input-output relation (\ref{inout}). Moreover, even for homodyne measurements one may design the experiment such that the phase fluctuations of the output field amplitude $\hat a_{\mathrm{out}}$ can be neglected
 (see Refs.~\cite{Elser}, \cite{Heim} for the corresponding experiment and Ref.~\cite{Toeppel} for its theoretical analysis).
In general, the atmospheric turbulence and the scattering effects may cause beam deformations,
speckles, etc., so  a multimode analysis of the transmitted light seems to be necessary.
However, the experiments under study  can be treated by an effective single-mode scenario (see
Appendixes~A and C in the Supplemental Material of Ref.~\cite{Vasylyev2012}).

Fine properties of fluctuating losses of the channel play a crucial role in free-space quantum communication.
Indeed, in many cases one postselects events with the large transmittance~\cite{Peuntinger, Semenov2010, Erven, Vallone2015a}.
In this context the widely-used log-normal distribution~\cite{ Milonni} may fail.
This  happens when fluctuations, which are related to the beam wandering~\cite{Vasylyev2012}, are significant.
In important practical situations, such as in the case of the considered atmospheric link of 1.6 km, both beam wandering and beam-spot distortion contribute in the PDT.
This situation can be properly described with the recently proposed elliptic-beam approximation~\cite{Vasylyev2016}.

In the present article we generalize  PDT based on the elliptic-beam approximation~\cite{Vasylyev2016}, in order to incorporate
the influence of random scatterers, such as haze particles and/or raindrops. There are two major effects of random scatterers on the beam: distortion of the beam shape including random deflection of its centroid and random losses due to scattering. We restrict our attention to the theoretical description of the former, while the latter effect is considered only phenomenologically. The theoretical PDTs are compared with the experimental distributions that were measured during  daytime and nighttime campaigns in Erlangen with an atmospheric link of $1.6$- km length. The nighttime measurements  were performed during the buildup of a hazy turbulent atmosphere, while the atmospheric daytime link was affected by light rainfall. The good agreement  between
the theoretical and experimental PDTs shows that our generalized elliptic-beam model is capable  of describing  quantum light propagation through the turbulent atmosphere even under diverse weather conditions.

The paper is organized as follows. In Sec.~\ref{sec:theory}  we discuss various theoretical aspects such as the input-output relations and the PDT model for atmospheric quantum channels. In Sec.~\ref{sec:experiment} the experimental setup is discussed and the experimental and theoretical PDTs are compared. The examples of the transfer of quadrature-squeezed and Gaussian-entangled light fields through the turbulent and scattering medium are presented in Sec.~\ref{sec:example}.  A summary  is given in Sec.~\ref{sec:conclusions}.

\section{Model of the turbulent and scattering medium }\label{sec:theory}

In the absence of absorption and scattering  the fluctuating losses on the receiver site arise mainly due to the
 finite aperture size of the receiving/detecting system.
Let us consider an initially Gaussian beam that propagates along the $z$~axis through the atmosphere. It impinges the circular aperture of radius $a$ placed at the distance $L$ from the transmitter. The atmospheric turbulence leads to random fluctuations of the beam shape and the beam-centroid position. As a result, the transmittance of such a beam through the aperture
\begin{align}\label{EtaDef}
 \eta=\int_{|\boldsymbol{\rho}|^2=a^2}\D^2\boldsymbol{\rho}\left|u(\boldsymbol{\rho},L)\right|^2,
\end{align}
is a fluctuating parameter cf. Ref.~\cite{Vasylyev2012,Vasylyev2016}. Here $u(\boldsymbol{\rho},L)$ is the beam envelope initially normalized in the transversal plane.

\subsection{Light beam in turbulent and scattering media}

\begin{figure}
 \includegraphics[width=0.45\textwidth]{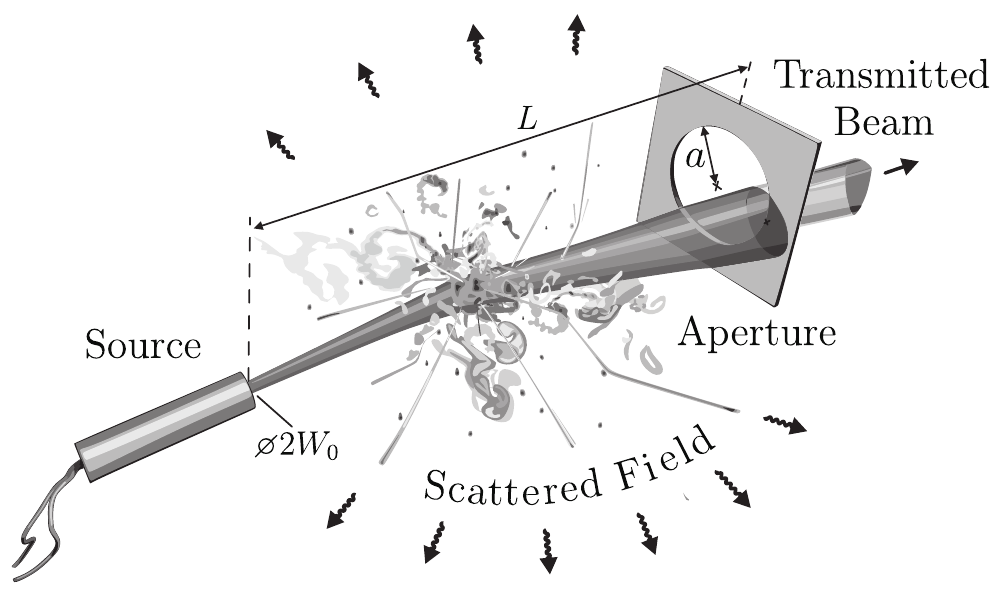}
 \caption{\label{fig:Setup}  The scheme of optical beam transmission through the turbulent and scattering atmosphere.  The beam with initial beam-spot radius $W_0$ undergoes random deformations and  deflections while propagating in the atmospheric link of length $L$.  A part of the radiation field is scattered and absorbed, which is the source of additional extinction losses. The transmitted beam is cut by the circular receiver aperture of radius $a$.  }
\end{figure}

The envelope function  $u(\boldsymbol{\rho},L)$ for an optical beam is obtained by solving the corresponding Helmholtz equation in the paraxial approximation \cite{Fante}. In this case the norm of the beam is preserved over the whole transmission path and the main effect comes from the losses  due to the finite aperture only. Such a technique gives a reasonable result in the case of turbulent atmosphere, when absorption and scattering effects can be neglected. However, these effects become essential for worse weather conditions in the presence of random scatterers, such as dust particles, aerosols, water droplets, etc. For the typical scenario, see Fig.~\ref{fig:Setup}.

In the presence of random scatterers we can still use Eq.~(\ref{EtaDef}) for determining the transmission efficiency. In this case, however, the norm of the beam envelope in the aperture plane can be less than one, due to scattering and absorption losses. While the absorption losses can be included in the paraxial approximation of the Helmholtz equation via the imaginary part of the permittivity, the consistent description of scattering losses requires  consideration beyond this approximation.

The resulting electromagnetic field in the scattering media is a superposition of two components: the transmitted beam and the scattered wave. The transmitted beam reaches the plane of the receiver aperture. The amplitude of the scattered wave is proportional to the scattering cross section, which determines the related losses \cite{IshimaruBook}. Since the norm of the whole electromagnetic wave should be preserved, the transmitted beam appears to be non-normalized.

In this paper we restrict our consideration to the transmitted-beam part of the electromagnetic wave. This part can still be described by using the paraxial approximation. The corresponding approach consistently describes distortions of the beam shape and deflections of the beam centroid by random scatterers. However, it does not yield the value of the corresponding scattering losses since we do not consider the noparaxial part of the scattered field. Hence we use a phenomenological approach for describing these losses.
For this purpose we additionally multiply the beam amplitude by the factor $\sqrt{\chi_{\rm ext}}$. The extinction factor $\chi_{\mathrm{ext}}{\in}[0,1]$ is a random variable that describes the absorption and scattering losses. The proper analysis of our experimental data (see Sec.~\ref{sec:experiment}) shows that the extinction factor can be considered as a nonfluctuating parameter, for the cases of rain and haze in the experimentally studied 1.6\,km link.

The solution of the paraxial wave equation together with the extinction factor in the form of a phase approximation of the Huygens-Kirchhoff method \cite{Mironov, Vasylyev2016} is given by
\begin{align}\label{uExpression}
 u(\boldsymbol{\rho},L)&=\sqrt{\chi_{\rm ext}}\int_{\mathbb{R}^2}\D^2\boldsymbol{\rho}^\prime u_0(\boldsymbol{\rho}^\prime) G(\boldsymbol{\rho},\boldsymbol{\rho}^\prime;L,0).
\end{align}
Here, the Gaussian beam envelope in Eq.~(\ref{uExpression}) at the transmitter plane $z{=}0$ reads 
\begin{align}\label{InitialU}
 u_0(\boldsymbol{\rho})=u(\boldsymbol{\rho},z{=}0)=\sqrt{\frac{2}{\pi W_0^2}}\exp\left[-\frac{1}{W_0^2}|\boldsymbol{\rho}|^2-\frac{ik}{2F}|\boldsymbol{\rho}|^2\right],
\end{align}
where $W_0$ is the beam-spot radius, $k$ is the optical wave number, and the beam is assumed to be focused  at  $z{=}F$.  The integral kernel in Eq.~(\ref{uExpression}) reads 
\begin{align}\label{IntKern}
 G(\boldsymbol{\rho},\boldsymbol{\rho}^\prime;z,z^\prime)&=\frac{k}{2\pi i(z-z^\prime)}\exp\left[\frac{i k|\boldsymbol{\rho}-\boldsymbol{\rho}^\prime|^2}{2(z-z^\prime)}\right]\nonumber\\
 &\times\exp\left[i S(\boldsymbol{\rho},\boldsymbol{\rho}^\prime;z,z^\prime)\right],
\end{align}
where
\begin{align}\label{RandPh}
 S(\boldsymbol{\rho},\boldsymbol{\rho}^\prime;z,z^\prime)=\frac{k}{2}\int\limits_{z^\prime}^{z}\D\xi\,  \delta \varepsilon\left(\boldsymbol{\rho}\frac{\xi-z^\prime}{z-z^\prime}+\boldsymbol{\rho}^\prime\frac{z-\xi}{z-z^\prime},\xi\right)
\end{align}
is the random phase contribution caused by the inhomogeneities of the fluctuating part of the real relative permittivity $\delta \varepsilon(\boldsymbol{\rho},z)$.  The random phase has contributions from random scattering, both at turbulent inhomogeneities and at additional random scatterers.

The statistical properties of the light field are affected by the statistical properties of the relative permittivity $\delta \varepsilon$. Note that the latter can be separated in two major contributions,
\begin{align}\label{EpsilonSplit}
\delta\varepsilon=\delta\varepsilon_\mathrm{turb}+\delta\varepsilon_\mathrm{scat},
\end{align}
with the parts related to the turbulence $\delta\varepsilon_\mathrm{turb}$ and to the random scatterers $\delta\varepsilon_\mathrm{scat}$. Here we  have assumed that the condition
$\delta\varepsilon_\mathrm{scat}\gg\delta\varepsilon_\mathrm{turb}$ applies. These two parts are also considered to be statistically independent. The correlation function for $\delta \varepsilon$ can be written as \cite{Fante}
 \begin{align}\label{PhStructureF}
\left\langle\delta\varepsilon(\boldsymbol{r}_1)\delta \varepsilon(\boldsymbol{r}_2)\right\rangle{=}\int\D^3\boldsymbol{K}\,\Phi_{\varepsilon}(\boldsymbol{K})\exp\left[i\boldsymbol{K}{\cdot}(\boldsymbol{r}_1-\boldsymbol{r}_2)\right],
  \end{align}
where $\boldsymbol{r}{=}(\boldsymbol{\rho}\quad z)^T$, $\boldsymbol{K}{=}(\boldsymbol{\kappa}\quad K_z)^T$ and $\Phi_{\varepsilon}(\boldsymbol{K})$ is the permittivity fluctuation spectrum, which splits as
\begin{align}\label{PhiSpectrum}
 \Phi_{\varepsilon}(\boldsymbol{K})=\Phi_{\varepsilon}^{\mathrm{turb}}(\boldsymbol{K})+\Phi_{\varepsilon}^{\mathrm{scat}}(\boldsymbol{K}),
\end{align}
due to the aforementioned statistical independence of $\delta\varepsilon_\mathrm{scat}$ and $\delta\varepsilon_\mathrm{turb}$. Using the Markov approximation \cite{IshimaruBook, Fante, Tatarskii}   we can further simplify Eq.~(\ref{PhStructureF}) and write
\begin{align}\label{SpectrumMarkov}
 \left\langle\delta\varepsilon(\boldsymbol{r}_1)\delta \varepsilon(\boldsymbol{r}_2)\right\rangle{=}2\pi\delta(z_1{-}z_2)\int\D^2\boldsymbol{\kappa}\Phi_\varepsilon(\boldsymbol{\kappa})e^{i\boldsymbol{\kappa}\cdot(\boldsymbol{\rho}_1{-}\boldsymbol{\rho}_2)},
\end{align}
where $\boldsymbol{\kappa}$ is the transverse wave vector.
The Markov approximation is well justified for the turbulent atmosphere \cite{Chumak}. For the scattering medium, this means that along the propagation direction the scattering on a certain particle is not influenced by the scattering on its neighbor particles~\cite{Zuev}.

For the turbulence part of the spectrum we use the Kolmogorov model~\cite{Tatarskii, Fante}
\begin{align}\label{PhiTurb}
 \Phi_{\varepsilon}^{\mathrm{turb}}(\boldsymbol{\kappa})=0.132 C_n^2 |\boldsymbol{\kappa}|^{-\frac{11}{3}},%\qquad \kappa_0{\ll}|\mathbf{k}|{\ll}\kappa_{\mathrm m},
\end{align}
which is defined in the inertial interval $|\boldsymbol{\kappa}|{\in}[\kappa_0,\kappa_{\mathrm m}]$ with $\kappa_0{\sim}1/L_0$ and $\kappa_{\mathrm m}{\sim}1/l_0$. Here $L_0$ and $l_0$ are the outer and inner scales of turbulence, respectively. The refractive index structure constant $C_n^2$
characterizes the strength of optical turbulence.  In the case of vertical or elevated links, such as ground-satellite links, the   dependence of structure constant on altitude should be taken into account.

 In Refs.~\cite{Lukin81, Yura} it was shown that the spectrum of the correlation function of fluctuations $\delta\varepsilon_{\mathrm{scat}}$ can be written for monodisperse scatterers of  size $\mathrm{d}_{\mathrm{scat}}$ as
 \begin{align}\label{ScatSp}
   \Phi_\varepsilon^{\mathrm{scat}}(\boldsymbol{\kappa})=\frac{2n_0 }{\pi k^4}|f_0(\boldsymbol{\kappa};\mathrm{d}_{\mathrm{scat}})|^2,
 \end{align}
 where $f_0(\boldsymbol{\kappa};\mathrm{d}_{\mathrm{scat}})$ is the amplitude of the wave scattered from a separate particle and $n_0$ is the mean number of scattering particles per unit volume.
The strict calculations of the  amplitude $f_0$ can be found in Refs~\cite{Hulst, IshimaruBook}. For scattering on haze or fog, one uses the Mie scattering theory, whereas the geometric scattering theory is applied for the  description of scattering on  drizzle and rain.

As shown in Refs.~\cite{Yura, Deepak, Lukin} one can approximate  $|f_0|$ in Eq.~(\ref{ScatSp}) in all aforementioned scattering scenarios by Gaussian functions. Equation~(\ref{ScatSp}) can be written  as
\begin{align}\label{PhiScat}
 \Phi_\varepsilon^{\mathrm{scat}}(\boldsymbol{\kappa})=\frac{n_0\zeta_0^4}{8\pi k^2} \exp\left[-\zeta_0^2|\mathbf{\boldsymbol{\kappa}}|^2\right].
\end{align}
This approximation means that the fluctuations of the scattering-related relative permittivity has a Gaussian  correlation function with the correlation length $2\zeta_0$ (see also~Appendix~\ref{app:PhaseApprox}). As was shown in Refs.~\cite{Yura, Deepak, Lukin}, the  parameter $\zeta_0$ is proportional to the particle size $\mathrm{d}_{\mathrm{scat}}$  and for the case of Mie scattering it depends additionally on the light wave number \cite{Deepak}. We also note that this model resembles
 the Gaussian phase screen model for the  correlation of phases due to random scattering \cite{Jakeman1974, Jakeman, Lukin}.

\subsection{Elliptic-beam model}

For many practical purposes we can restrict the effect of the atmosphere on the beam shape to elliptic deformations only. It is also important to include  random wandering of the beam centroid into consideration. This is the main idea behind the recently proposed method of the elliptic-beam approximation~\cite{Vasylyev2016}.

In this model, the PDT, which appears in Eq.~(\ref{Pinout}), is given by
\begin{align}\label{PDTeq}
 \mathcal{P}(\eta){=}\frac{2}{\pi}\int_{\mathbb{R}^4}\D^4\mathbf{v}\int_0^{\pi/2}\D\varphi\,\rho_G(\mathbf{v};\boldsymbol{\mu},\boldsymbol{\Sigma})\delta\left(\eta-\eta(\mathbf{v},\varphi)\right).
\end{align}
Here $\eta(\mathbf{v},\varphi)$ is defined by Eq.~(\ref{EtaDef}) and is specified for the elliptic beam. Its explicit form is given by Eq.~(\ref{etaElBeam}) in Appendix~\ref{app:EllModel}. This is a function of the random vector $\mathbf{v}{=}(x_0\quad y_0\quad \Theta_1\quad \Theta_2)^T$ and the angle $\varphi$ related to the beam-ellipse orientation. The parameters $x_0$ and $y_0$ correspond to the beam-centroid coordinates and the parameters $\Theta_i{=}\ln(W_i^2/W_0^2)$, $i{=}1,2$, correspond to  the ellipse semiaxis $W_i$. The distribution function for these parameters, $\rho_G(\mathbf{v};\boldsymbol{\mu},\boldsymbol{\Sigma})$, is assumed to be Gaussian with the vector of mean values $\boldsymbol{\mu}$ and the covariance matrix $\boldsymbol{\Sigma}$.   The knowledge of parameters $\boldsymbol{\mu}$ and $\boldsymbol{\Sigma}$ enables one to evaluate numerically the integral (\ref{PDTeq}) and therefore to estimate the PDT (for details see Appendix~\ref{app:PDT}).

According to the procedure, which is discussed in Ref.~\cite{Vasylyev2016}, we calculate the values of $\boldsymbol{\mu}$ and $\boldsymbol{\Sigma}$ for the focused beam $L{=}F$ for details see Appendixes~\ref{app:PhaseApprox} and~\ref{app:EllModelParam}. Unlike the consideration in Ref.~\cite{Vasylyev2016}, here we also take into account the contributions from the spectrum of random scatterers. Specifically, it turns out that these scatterers do not affect the beam wandering, such that the diagonal elements of the matrix $\boldsymbol{\Sigma}$ related to the parameters $x_0$ and $y_0$ are given by
\begin{align}
 \langle x_0^2\rangle=\langle y_0^2\rangle=0.33 W_0^2\sigma_R^2\Omega^{-\frac{7}{6}},\label{BW}
\end{align}
where  $\Omega=\frac{k W_0^2}{2L}$ is the Fresnel number and
\begin{align}\label{Rytov}
 \sigma_R^2=1.23 C_n^2 k^{\frac{7}{6}}L^{\frac{11}{6}}
\end{align}
is the Rytov variance \cite{Fante, IshimaruBook} which characterizes the strength of phase fluctuations of transmitted light due to scattering on turbulent inhomogeneities.
 This effect can be explained by the fact that the  random scatterers are much smaller than the beam diameter.

The part of the mean vector and the covariance matrix related to the parameters $\Theta_i$ are obtained via mean values and the (co)variances of the squared ellipse semiaxes $W_i^2$,
\begin{align}\label{ThetaMean}
 \langle\Theta_i\rangle=\ln\left[\frac{\langle W_i^2\rangle}{W_0^2\left(1+\frac{\langle(\Delta W_i^2)^2\rangle}{\langle W_i^2\rangle^2}\right)^{1/2}}\right],
\end{align}
\begin{align}\label{ThetaVariance}
 \langle\Delta\Theta_i\Delta\Theta_j\rangle=\ln\left(1+\frac{\langle\Delta W_i^2\Delta W_j^2\rangle}{\langle W_i^2\rangle\langle W_j^2\rangle}\right).
\end{align}
These parameters are calculated with the aforementioned technique and read 
\begin{align}
\langle W_{1/2}^2\rangle=\frac{W_0^2}{\Omega^2}\Bigl[1+\Xi+2.96\sigma_R^2\Omega^{\frac{5}{9}}\Bigr],\label{W2mean}
\end{align}
\begin{align}
 \langle \Delta W_i^2\Delta W_j^2\rangle{=}(2\delta_{ij}{-}0.8) \frac{W_0^4}{\Omega^{\frac{19}{6}}}\left[1+\Xi\right]\sigma_R^2,\label{W2covar}
\end{align}
where
\begin{align}\label{divergence}
 \Xi=\frac{2}{3}\sigma_{S,\mathrm{scat}}^{2}\frac{W_0^2}{4\zeta_0^2}.
\end{align}
is the beam divergence parameter due to the random scattering. In Eq.~(\ref{divergence}) we have introduced the phase variance of transmitted light due to random scattering as
\begin{align}\label{PhVar}
 \sigma_{S,\mathrm{scat}}^{2}=\frac{\pi}{4}n_0 L\zeta_0^2.
\end{align}
Thus,  the random scattering contributes to the beam broadening and to the (co)variances of the beam deformation fluctuations. It is also worth  mentioning that in the considered approximation the parameters $\Theta_i$ are statistically independent   from the beam centroid coordinates $x_0$ and $y_0$ such that the corresponding covariances vanish (for more details see the Supplemental Material of Ref.~\cite{Vallone2016}).

The beam broadening term due to the random scattering in Eq.~(\ref{W2mean}) can be compared with the results obtained within the  small-angle approximation of the radiative transfer equation~\cite{Tam, Box, Ishimaru}
\begin{align}\label{Transport}
 \langle W^2_{\rm scat}\rangle=\frac{W_0^2}{\Omega^2}+\frac{2}{3} A \tau L^2\overline{\psi^2},
\end{align}
where the first term represents beam broadening in free space and the second term is due to random scattering. Here $A$ is the albedo of a single scatterer, i.e., the ratio of the scattering cross section $\sigma_{\rm scat}$ to the extinction cross section $\sigma_{\rm ext}$, $\tau{=}-\ln\chi_{\rm ext}$ is the unitless optical distance~\cite{Ishimaru}, and $\overline{\psi^2}$ is the mean square of the scattering angle~\cite{Hulst, Box}. Comparing Eq.~(\ref{Transport}) with the first two terms in (\ref{W2mean}),  we obtain the  expression for the beam divergence parameter
\begin{align}
\Xi=\frac{2}{3}\frac{\Omega^2}{W_0^2} A\tau\overline{\psi^2}.
\end{align}
This equation relates the model parameters $\zeta_0$ and $\sigma_{S,\mathrm{scat}}^2$ to the  properties of scattering media.

\section{Experimental results and discussion}\label{sec:experiment}

We apply the theory to experimental data collected during nighttime and daytime campaigns for various weather conditions.
The experimental data were collected in experiments on the free-space distribution of squeezed states of light in an urban free-space channel of 1.6-km length in Erlangen \cite{Peuntinger, Usenko, Heim2014}. During the measurements, the relative transmission is recorded, which can also be used for atmospheric studies, as it is the case here.
The experimental setup is divided into a sender and a receiver. The former consists of a laser (Origami, Onefive GmbH) with central wavelength at 1559-nm emitting pulses  with lengths of 200 fs at a repetition rate of 80 MHz. After this light is frequency doubled in a periodically poled lithium niobate crystal (MSHG1550-0.5-0.3, Covesion Ltd.), it is guided through a polarization-maintaining photonic crystal fiber, thereby generating polarization-squeezed states of light that are sent to the receiver. More details on the sender may be found in~\cite{Peuntinger}.

\begin{figure}[ht]
 \includegraphics[width=0.45\textwidth]{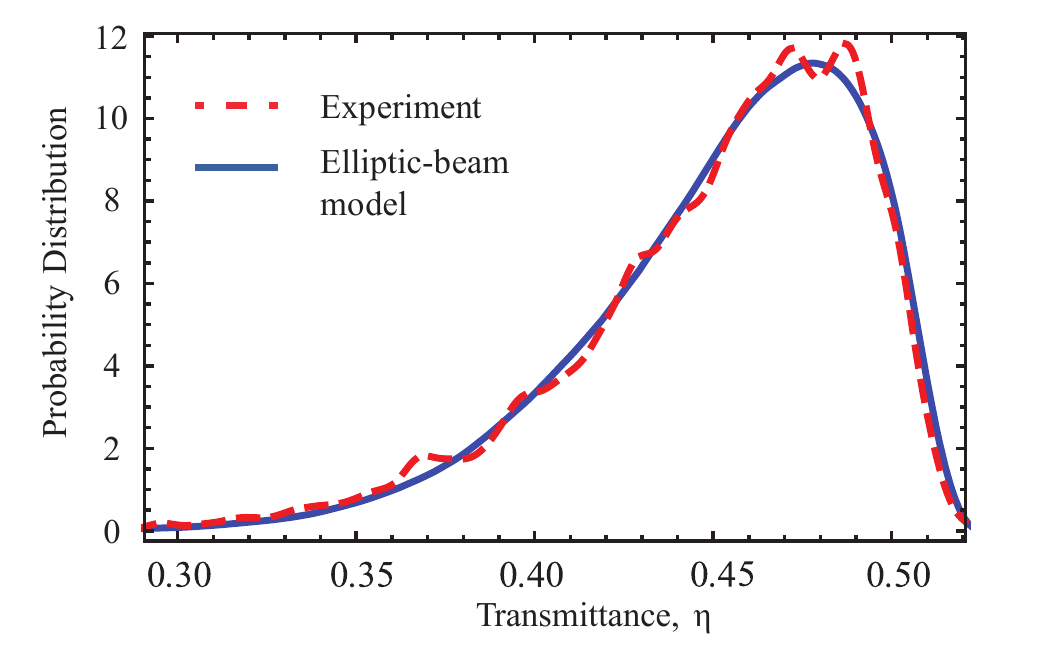}
 \caption{\label{fig:Night1} (Color online)  Elliptic-beam approximation PDT (solid line)  compared with the experimental PDT (dashed line). The measurement was performed at night  on 24. August 2016 at 00:20 (local time); the data acquisition time is $12$ s. The other parameters are a wavelength of $780$ nm, the  initial spot radius $W_0=20$ mm, the aperture radius $a{=}75$ mm, the Rytov parameter $\sigma_R^2{=}1.78$,  the beam divergence parameter due to random scattering $\Xi{=}5$, and the extinction factor  $\chi_{\mathrm{ext}}{=}0.51$.   The last three parameters are derived from the fitting procedure  of the theoretical PDT to the experimental data. }
\end{figure}

\begin{figure}[ht]
 \includegraphics[width=0.45\textwidth]{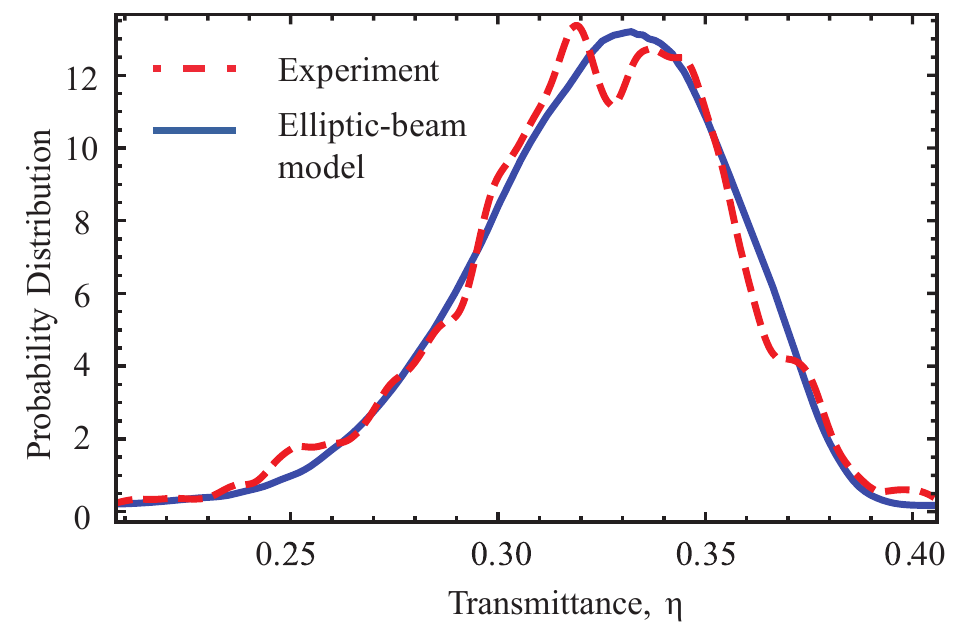}
 \caption{\label{fig:Night2} (Color online)  Elliptic-beam approximation PDT (solid line)  compared with the experimental PDT (dashed line). The measurement was performed at night on 24. August 2016 at 02:00 (local time); the data acquisition time is $9$ s. The estimated  from the fit Rytov parameter is $\sigma_R^2{=}1.05$,  the beam divergence parameter  $\Xi{=}12$, and  $\chi_{\mathrm{ext}}{=}0.40$.}
\end{figure}

\begin{figure}[ht]
 \includegraphics[width=0.45\textwidth]{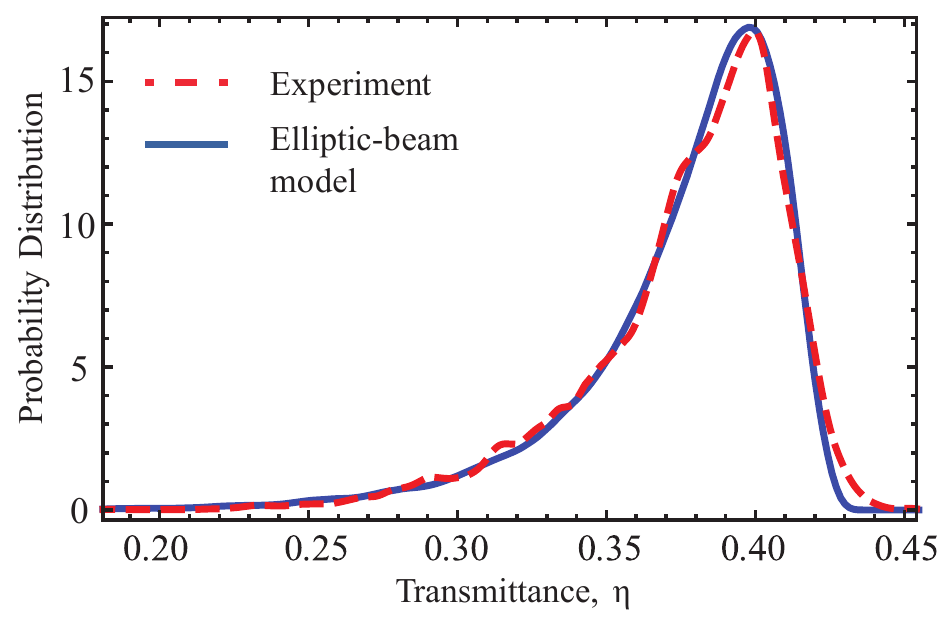}
 \caption{\label{fig:PDTrain}  (Color online) Experimental (dashed line) and theoretically fitted (solid line) PDTs  for atmospheric quantum channel in the presence of rainfall.  The measurement was performed at daytime on 08. June 2016 at 11:15 (local time); the data acquisition time is $14$ s. The estimated Rytov parameter $\sigma_R^2=2.88$, the beam divergence parameter due to scattering on haze $\Xi{=}0.2$, the path-averaged rain intensity  $\overline{\mathcal{I}}{=}3.2\text{mm}/\text{h}$, and $\chi_{\mathrm{ext}}{=}0.43$.}
\end{figure}

At the receiver the light is collected by an achromatic lens with a diameter of 150-mm and a focal length
of ${f{=}800\,\text{mm}}$. The beam is split into two beams by a 50-50 or 90-10 beam splitter for the nighttime
measurements and daytime measurement presented here, respectively.
Each of the beams is led to a
Stokes detection setup consisting of a half waveplate, a polarization beam splitter, and two custom-made
pin-diode (S3883, quantum efficiency $\eta_{\mathrm{det}}{=}0.9$, Hamamatsu Photonics K.K.) detectors. The dc output of
these detectors is proportional to the intensity of the impinging light and is sampled with a rate of
80-kHz.
The typical data acquisition time for the PDTs is on the order of several seconds. The exact durations for the shown data can be found in the captions of Figs.~\ref{fig:Night1}, \ref{fig:Night2}, and
\ref{fig:PDTrain}, respectively.
Comparing the sum of the four detector outputs with the sent laser power gives the
transmission for each sample. The constant losses due to imperfections of all the
optical elements in the setup add up to a value of $\eta_{\mathrm{opt}}{=}0.88$.

The nighttime measurements on 24. August 2016  were performed   at $00{:}20$ and at $02{:}00$ of local time. During two hours with separate measurements the temperature dropped from 16$^{\circ}\mathrm{C}$ to 14$^{\circ}\mathrm{C}$, whereas the relative humidity increased from $85\%$ to  $94\%$. The increase of humidity has led to the  formation of more dense haze and as a consequence to the reduction of the optical visibility.  The corresponding experimental PDTs are obtained by the smooth kernel method~\cite{Wand} and are shown in Figs.~\ref{fig:Night1} and~\ref{fig:Night2}.

The theoretical curves in Figs.~\ref{fig:Night1} and \ref{fig:Night2} were calculated by using the elliptic-beam approximation for the PDT [cf.~Eq.~(\ref{PDTeq})] by using Monte Carlo methods. We estimated the value for the Rytov parameter $\sigma_R^2$,  the  divergence parameter $\Xi$, and the extinction factor $\chi_{\mathrm{ext}}$ with the Pierson's $\chi^2$ criterion~\cite{Agresti}.  The increase of humidity between the two  measurements led to the increase of the haze particle sizes  and the haze number density. The decrease of temperature, on the other hand, reduced the intensity of thermal fluctuations in the atmosphere and hence the strength of optical   turbulence. These effects cause  a significant change of transmittance statistics. Comparing Fig.~\ref{fig:Night1} with Fig.~\ref{fig:Night2}, we see that with an increase of the divergence parameter the PDT becomes more symmetric and resembles a Gaussian distribution. Therefore, the channel
presented in Fig.~\ref{fig:Night2} has a smaller
probability of attaining high  values of transmittance in comparison to the channel in Fig.~\ref{fig:Night1}.

\begin{figure}[h!]
 \includegraphics[width=0.45\textwidth]{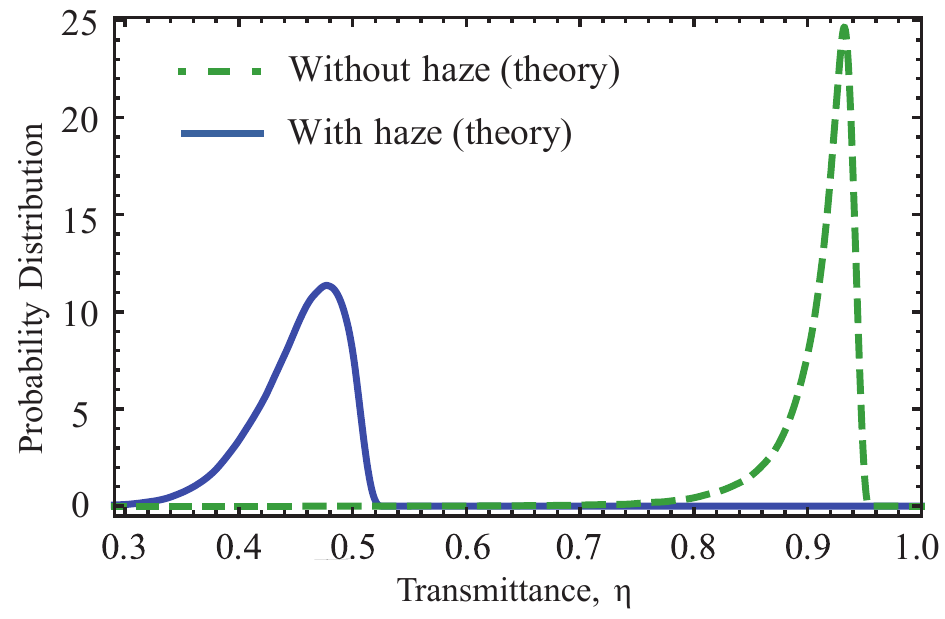}
 \caption{\label{fig:HazeEffect} (Color online) Influence of haze on transmission characteristics of the atmospheric channel. The PDT with the inclusion of random scattering due to haze (solid line) is calculated for the same parameters as in Fig.~\ref{fig:Night1}, i.e., it corresponds to the experimental data.  The PDT without inclusion of haze (dashed line) is  theoretically deduced; extinction losses are $\chi_{\mathrm{ext}}{=}0.94$ due to molecular absorption.  }
\end{figure}

 In order to demonstrate the impact of haze, in Fig.~\ref{fig:HazeEffect} we show the theoretical PDT from Fig.~\ref{fig:Night1} (solid line) and compare it to the theoretically one deduced when there would not be any haze (dashed line). The extinction of the light signal due to haze  shifts the distribution to the smaller values of the transmittance. At the same time, the random scattering broadens the distribution.

\begin{figure}[ht]
 \includegraphics[width=0.45\textwidth]{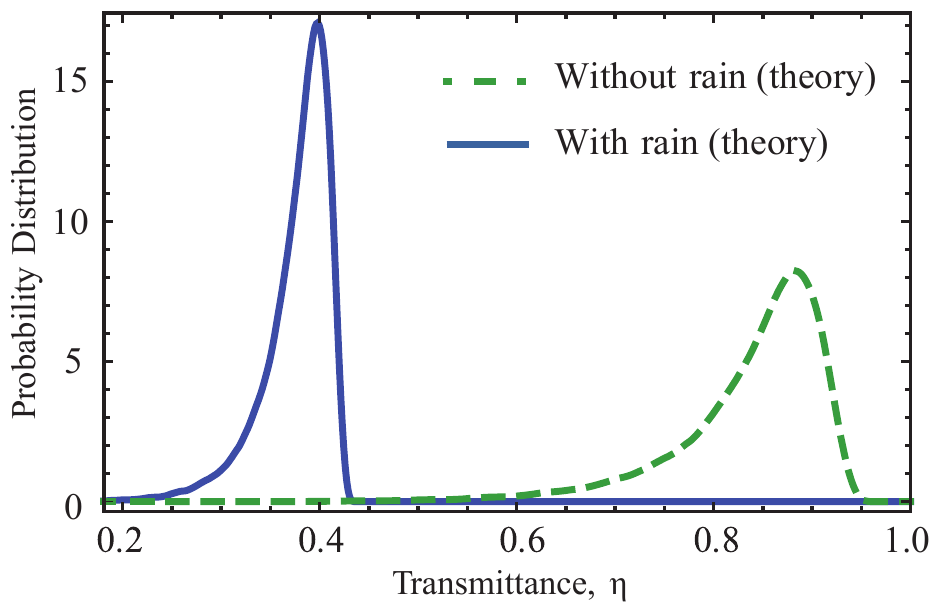}
 \caption{\label{fig:RainEffect}   (Color online)  Atmospheric channel transmittance distribution with (solid line) and without (dashed line) light rain.  The latter PDT includes additional extinction losses  $\chi_{\mathrm{ext}}{=}0.94$ due to molecular absorption. The solid line corresponds to the elliptic-beam model
 fit in Fig.~\ref{fig:PDTrain}, whereas the dashed line is theoretically
 deduced. }
\end{figure}

The daytime measurement  was performed  on 08. June 2016 at 11:15 (local time). The meteorological data for this day were the following:  $76\%$  humidity, a temperature of $17.6 ^{\circ}\mathrm{C}$,  $3\,\text{mm}/\text{h}$ of mean rain rate during the day. Figure~\ref{fig:PDTrain} shows the experimental (dashed line) and the theoretical (solid line) PDTs. During the day measurements the atmospheric channel  is characterized by a larger value of the Rytov parameter ($\sigma_R^2=2.88$) in comparison to the night measurements, because of increased turbulence due to thermal convection and wind shear. We have observed that  light rain introduces a minor contribution to the phase fluctuation given by Eq.~(\ref{RandPh}) and estimated the divergence parameter as $\Xi=0.2$, i.e., the beam broadening and beam deformation due to rainfall is small.  The more pronounced effect of the rainfall is connected with the   contribution to the extinction factor $\chi_{\rm ext}$. We have used an empirical formula \cite{Zuev, 
Rainfall} that connects the
extinction factor with the path-averaged rainfall intensity $\overline{\mathcal{I}}$ (in mm/h), and propagation distance $L$ (in m) as
 \begin{align}\label{RainExtinction}
  \chi_{\rm ext}=\exp\bigl[-210\, \overline{\mathcal{I}}^{0.74}L\bigr].
 \end{align}
The dependence of the extinction coefficient on the water content of precipitations is given in Ref.~\cite{Mori}.
In analogy to Fig.~\ref{fig:HazeEffect},
Fig.~\ref{fig:RainEffect} compares the theoretical PDTs with (solid line) and without (dashed line) inclusion of losses due to scattering and absorption by rainfall. One can see that the PDT is shifted to lower values of transmittance and it is narrowed.

\section{Quadrature squeezing and Gaussian entanglement} \label{sec:example}

In order to illustrate the capability of atmospheric channels to preserve nonclassical properties of quantum light, we consider the propagation of quadrature squeezed and Gaussian entangled states through turbulence, rain, and haze. The knowledge of the PDT (\ref{PDTeq}) allows us to analyze the quantum properties of propagating light with the help of the input-output  relation (\ref{Pinout}). Alternatively, one can use directly the input-output relation (\ref{inout}) and the PDT is used for the calculation of moments $\langle\sqrt{\eta}\rangle$, $\langle\eta\rangle$, etc.

\begin{figure}[ht]
 \includegraphics[width=0.45\textwidth]{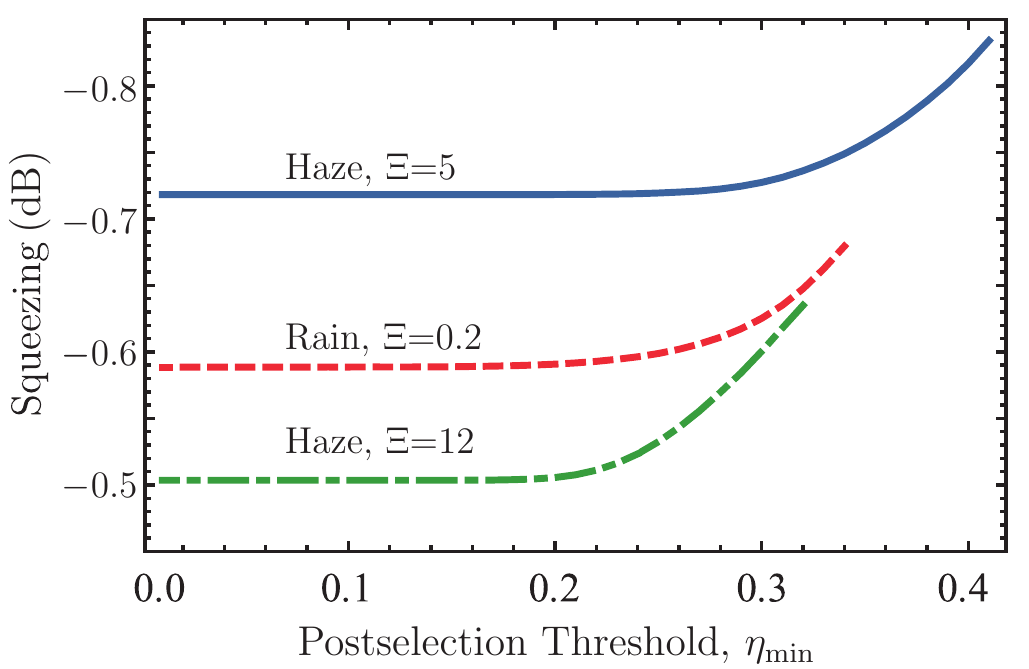}
 \caption{\label{fig:Squeezing}  (Color online) Transmitted value of squeezing as a function of postselection threshold $\eta_{\rm min}$. The input light is squeezed to $-2.4$ dB and sent through  three atmospheric channels presented in  Figs.~\ref{fig:Night1},   \ref{fig:Night2}, \ref{fig:PDTrain}. The predicted behaviors are theoretically evaluated on the basis of the experimental PDTs.  The solid line corresponds to the nighttime channel with the Rytov parameter $\sigma_R^2{=}1.78$, divergence parameter $\Xi{=}5$,  extinction losses $\chi_{\mathrm{ext}}{=}0.51$ and mean transmittance $\langle\eta\rangle{=}0.36$. The dashed  line corresponds to the rainy daytime  channel with $\sigma_R^2{=}2.88$, $\Xi{=}0.2$,    $\chi_{\mathrm{ext}}{=}0.43$, and $\langle\eta\rangle{=}0.29$. The dash-dotted line corresponds to the nighttime channel with haze, $\sigma_R^2{=}1.05$, $\Xi{=}12$, $\chi_{\mathrm{ext}}{=}0.40$, and $\langle\eta\rangle{=}0.26$. For all curves the additional losses on optical components $\eta_{\
mathrm{opt}}{=}0.88$ and detection efficiency $\eta_{\mathrm{det}}{=}0.9$ are included. All
other parameters are the same as in Fig.~\ref{fig:Night1}.  }
\end{figure}

The propagation of squeezed light through the turbulent atmosphere has been studied both theoretically~\cite{Vasylyev2012} and experimentally~\cite{Peuntinger}. It has been shown that the postselection procedure of transmission events with  transmittance values greater than the postselection threshold $\eta_{\mathrm{min}}$ yields larger values of the transmitted squeezing  \cite{Peuntinger, Heersink}.
We consider the propagation of quadrature squeezed light ($-2.4$ dB) at $\lambda{=}780$\,nm over 1.6\,km under different atmospheric conditions.  Figure~\ref{fig:Squeezing} shows the values of squeezing as a function of the postselection threshold for atmospheric channels. The corresponding PDTs are shown in Figs.~\ref{fig:Night1}, \ref{fig:Night2}, and \ref{fig:PDTrain}. The values of transmitted squeezing as well as the maximal postselection threshold values depend on the mean transmittance $\langle\eta\rangle$. The most favorable conditions for squeezing transmission  (see the solid line in Fig.~\ref{fig:Squeezing}) are for the night-time
measurement with haze and a low value of the divergence parameter $\Xi$  [cf.~Eq.~(\ref{divergence})].  The stronger optical turbulence during the daytime transmission  diminishes the detectable squeezing value (see the dashed line in Fig.~\ref{fig:Squeezing}). At the same time the beam divergence due to scattering plays
 a minor role here. The dash-dotted line in Fig.~\ref{fig:Squeezing} shows that the presence of  denser haze during the second nighttime measurement contributes to a stronger beam divergence and hence to a smaller efficiency of squeezed light transmission.

As the next example we consider the transmission of Gaussian entanglement of a two-mode squeezed vacuum (TMSV) state in the turbulent atmosphere with haze or rain. Here we  closely follow  the theoretical analysis of  Ref.~\cite{Bohmann}, where it was shown that, in contrast to the channels with deterministic losses, the propagation through the atmosphere with fluctuating transmittance yields certain restrictions on the squeezing degree of the TMSV.

We consider the scenario when one mode of the entangled light fields is sent through the atmospheric channel (field mode $A$), whereas the second one is analyzed locally at the transmitter site (field mode $B$). For the
transmitted and detected state we apply the Simon entanglement criterion \cite{Simon} in the form  found in Ref.~\cite{Shchukin}, stating that any two-mode Gaussian state is entangled if and only if
\begin{align}\label{Simon}
 \mathcal{W}=\det V^{\mathrm{PT}}<0,
\end{align}
where $V^{\mathrm{PT}}$ is the partial transposition of the matrix
\begin{align}
 V=\left(\begin{array}{c c c c}
          \langle\Delta\hat a^\dagger\Delta\hat a\rangle& \langle\Delta\hat a^{\dagger 2}\rangle& \langle\Delta\hat a^\dagger\Delta\hat b\rangle&\langle\Delta\hat a^\dagger\Delta\hat b^\dagger\rangle\\
          \langle\Delta\hat a^2\rangle&\langle\Delta\hat a\Delta\hat a^\dagger\rangle&\langle\Delta\hat a\Delta\hat b\rangle&\langle\Delta\hat a\Delta\hat b^\dagger\rangle\\
          \langle\Delta\hat a\Delta\hat b^\dagger\rangle&\langle\Delta\hat a^\dagger\Delta\hat b^\dagger\rangle&\langle\Delta\hat b^\dagger\Delta\hat b\rangle& \langle\Delta\hat b^{\dagger 2}\rangle\\
          \langle\Delta\hat a\Delta\hat b\rangle&\langle\Delta\hat a^\dagger\Delta\hat b\rangle&\langle\Delta\hat b^2\rangle& \langle\Delta\hat b\Delta\hat b^{\dagger}\rangle
         \end{array}
\right).
\end{align}
The  matrix $V$ is the second-order matrix of moments of the bosonic creation and annihilation operators of the field modes $A$ and $B$,  where $\Delta\hat x=\hat x-\langle \hat x\rangle$ with $\hat x=\hat a,\hat b$. The Simon entanglement test $\mathcal{W}_{\rm atm}$ for the state when one mode is transmitted through the atmosphere is obtained by applying the input-output relation (\ref{inout}) to the field mode $A$, i.e., for the operators $\hat a$ and $\hat a^\dagger$. As it was shown in Ref.~\cite{Bohmann}, the entanglement test $\mathcal{W}_{\mathrm{atm}}$ for fluctuating loss channels contains a term that depends on the coherent displacement $|\langle\hat a\rangle|$. This feature yields some restrictions on the value of the coherent displacement  since for some boundary value  $|\langle\hat a\rangle|$ the Simon test becomes positive. Similarly, there exists some boundary value of the squeezing parameter $\xi$ above which the Gaussian entanglement is not preserved.

\begin{figure}[ht]
 \includegraphics[width=0.45\textwidth]{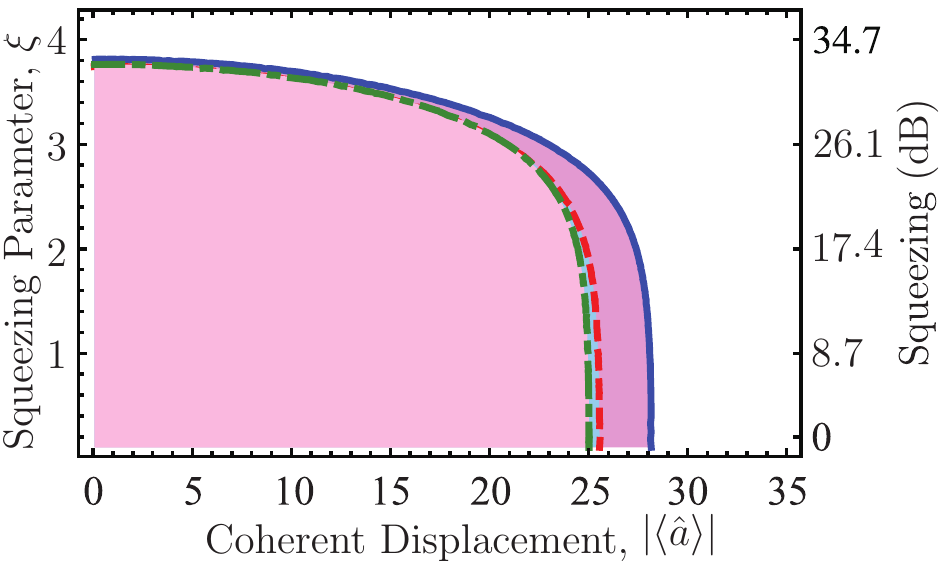}
 \caption{\label{fig:Entangl}(Color online) Shaded areas represent the regions where the entanglement can be  verified by applying the Simon entanglement test, which is a  function of the squeezing parameter $\xi$ and of the coherent displacement  $|\langle\hat a\rangle|$. The solid, dashed, and dash-dotted lines  correspond to the bounds of the  region where entanglement survives, for  the respective channels listed in the caption of Fig.~\ref{fig:Squeezing}. }
\end{figure}

In Fig.~\ref{fig:Entangl} we show the regions where the Gaussian entanglement can be verified for the three atmospheric channels characterized by the PDTs given in Figs.~\ref{fig:Night1}, \ref{fig:Night2}, and \ref{fig:PDTrain}. The boundary values of the coherent amplitude  $|\langle\hat a\rangle|$ and squeezing parameter $\xi$ are shown by solid, dashed, and dash-dotted lines for the corresponding channels. The boundary values of the squeezing parameter lie beyond the experimentally  obtainable  squeezing strengths that were obtainable in our experiment. However for long propagation paths this border could be reached already for practically generated TMSV states. This effect therefore should be  taken into account when applying TMSV-based quantum protocols for long-distance quantum communication.  In Fig.~\ref{fig:Entangl}  we also see that the presence of random scattering by haze particles shrinks the area  where the
Gaussian entanglement   persists, by reducing the boundary value of the coherent displacement amplitude. Similarly  to
the case of quadrature squeezing transmission, the particular daytime channel with rain preserves Gaussian entanglement better than the hazy nighttime channels.

\section{Summary and Conclusions} \label{sec:conclusions}
A quantum state that is transmitted through an atmospheric quantum link experiences fluctuating losses that can spoil or completely destroy its nonclassical properties. Here we studied a realistic intracity free-space quantum channel that has turbulence- and scattering-induced fluctuating losses. Our experimental results show that the transmittance  statistics for Gaussian beams strongly depends  on the meteorological conditions and can change drastically  within a few hours between two measurements. Our theoretical studies explained this situation by taking into account not only the atmospheric turbulence but also  the random scattering on haze particles or on  raindrops. Using the elliptic-beam model for the beam transmitted through the atmosphere and impinging on the receiver aperture, we have shown that random scattering on haze particles  contributes to the beam broadening and beam shape deformation. The action of rain shows minor beam broadening and deformation effects, but
it contributes to the extinction losses.

We have studied the transmission of  quadrature squeezing and Gaussian entanglement through realistic quantum optical links with turbulence, haze and rain. We have found that  a detectable squeezing value depends  on the propagation conditions and it is strongly affected by random scattering. For example, the daytime transmission in rain preserves  squeezing better than the nighttime  transmission in haze, despite the fact that the optical turbulence is considerably stronger during the day. Similar effects have been found by analyzing the transmission of Gaussian entanglement  through atmosphere.  Random scattering on haze particles  constricts the area  of the values  of the squeezing parameter and coherent amplitude, for which  entanglement is verified. The obtained results may be useful for the analysis of quantum communication protocols in intracity atmospheric channels under diverse weather and
day-time conditions.

\acknowledgments The authors are grateful to M. Bohmann for useful and enlightening
discussions. The work was supported by the Deutsche Forschungsgemeinschaft through
Project No. VO 501/21-2. The authors thank  G. Leuchs for enlightening discussions and  our
colleagues at the FAU computer science building for their kind support and for hosting the receiver
station.

\appendix
\section{Aperture transmittance}\label{app:EllModel}

In this appendix we remind the reader of some details on the elliptic-beam model for the PDT \cite{Vasylyev2016}.  We choose the coordinate system  such that the $z$ axis is aligned along the line
that connects the centers of transmitter and receiver apertures. The distance between the transmitter and the receiver aperture plane is $z{=}L$.
The transmission efficiency of an elliptic beam through  a circular aperture of radius $a$  is given by Eq.~(\ref{EtaDef}), where the beam intensity at the aperture plane is assumed to have the Gaussian form
\begin{align}\label{IntElip}
 |u(\boldsymbol{\rho},L)|^2{=}\frac{2\chi_{\mathrm{ext}}}{\pi\sqrt{\det\mathbf{S}}}\exp\Bigl[-2(\boldsymbol{\rho}{-}\boldsymbol{\rho}_0)^T\mathbf{S}^{-1}(\boldsymbol{\rho}{-}\boldsymbol{\rho}_0)\Bigr].
\end{align}
Here, $\boldsymbol{\rho}{=}(x\quad y)^T$ is the transverse coordinate, $\boldsymbol{\rho}_0{=}(\rho_0\cos\phi_0\quad \rho_0\sin\phi_0)^T$ is the beam centroid position coordinate, $\mathbf{S}$
is the real, symmetric, positive-definite spot-shape matrix,  and $\chi_{\rm ext}$ is the extinction factor due to absorption and scattering. In general, the intensity   (\ref{IntElip}) has an elliptic profile.
Applying the rotation   by  a certain  angle $\varphi$, we can bring
the spot shape matrix into diagonal form with the elements $W_{i}^2$, $i{=}1,2$, which are the squared major semi-axes of the ellipse.

Substituting Eq.~(\ref{IntElip}) in Eq.~(\ref{EtaDef}), one can show that the transmission efficiency
 can be approximated by the following expression  (cf.~Ref.~\cite{Vasylyev2016}):
\begin{align}\label{etaElBeam}
 \eta=\eta_0\exp\left\{-\left[\frac{\rho_0/a}{R\left(\frac{2}{W_{\rm eff}(\varphi-\phi_0)}\right)}\right]^{\lambda\left(\frac{2}{W_{\rm eff}(\varphi-\phi_0)}\right)}\right\}.
\end{align}
Here the maximal transmittance for a centered beam,
\begin{align}
 \eta_0&=1{-}\I_0\left(a^2\frac{W_1^2-W_2^2}{W_1^2W_2^2}\right)e^{-a^2\frac{W_1^2+W_2^2}{W_1^2W_2^2}}\nonumber\\
 &-2\Bigl[1{-}e^{-\frac{a^2}{2}\Bigl(\frac{1}{W_1}{-}\frac{1}{W_2}\Bigr)}\Bigr]\\
 &\times\exp\left[-\left\{\frac{\frac{(W_1+W_2)^2}{|W_1^2-W_2^2|}}{R\left(\frac{1}{W_1}-\frac{1}{W_2}\right)}\right\}^{\lambda\left(\frac{1}{W_1}-\frac{1}{W_2}\right)}\right]\nonumber
\end{align}
is a function of the two eigenvalues $W_{i}^2$ of the spot-shape matrix $\mathbf{S}$ and $\I_n(x)$ is the modified Bessel function of $n$-th order.
The shape $\lambda$ and scale $R$ functions are given by
\begin{align}
 \lambda(\xi)&=2a^2\xi^2\frac{e^{-a^2\xi^2}\I_1\left(a^2\xi^2\right)}{1-\exp\left[-a^2\xi^2\right]\I_0\left(a^2\xi^2\right)}\nonumber\\
 &\times\left[\ln\left(2\frac{1-\exp\left[-\frac{1}{2}a^2\xi^2\right]}{1-\exp\left[-a^2\xi^2\right]\I_0\left(a^2\xi^2\right)}\right)\right]^{-1},
\end{align}
\begin{align}
 R(\xi)=\left[\ln\left(2\frac{1-\exp\left[-\frac{1}{2}a^2\xi^2\right]}{1-\exp\left[-a^2\xi^2\right]\I_0\left(a^2\xi^2\right)}\right)\right]^{-\frac{1}{\lambda(\xi)}},
\end{align}
where the effective squared spot radius
\begin{align}
& W_{\rm eff}^2(\varphi-\phi_0)=4a^2\Bigl[\mathcal{W}\Bigl(\frac{4a^2}{W_1W_2}e^{2a^2\left(\frac{1}{W_1^2}+\frac{1}{W_2^2}\right)}\Bigr.\Bigr.\\
 &\qquad\times\Bigl.\Bigl. e^{a^2\left(\frac{1}{W_1^2}-\frac{1}{W_2^2}\right)\cos(2\varphi-2\phi_0)}\Bigr)\Bigr]^{-1}
\end{align}
is expressed with the help of  the Lambert  function  $\mathcal{W}(x)$ (cf. Ref.~\cite{Corless}). Thus, the elliptic beam transmittance (\ref{etaElBeam}) is a function of the random variables $\boldsymbol{\rho}_0=(x_0\quad y_0)^T$, $W_1^2$, $W_2^2$, and $\phi{=}\varphi-\phi_0$; or, alternatively, of the variables $\boldsymbol{\rho}_0$, $\Theta_1$, $\Theta_2$, and $\phi$, where
\begin{align}\label{Theta}
 W_i^2=W_0^2\exp{\Theta_i},
\end{align}
with $W_0^2$ being the beam spot radius at the transmitter.

\section{Evaluation of the PDT}\label{app:PDT}

In this appendix we discuss how to numerically evaluate the PDT in Eq.~(\ref{PDTeq}), based on the knowledge of relevant atmospheric and beam parameters. To this end one should proceed with the following steps.
\begin{enumerate}[label=(\roman*)]
 \item One calculates the components of covariance matrix $\boldsymbol{\Sigma}$ and mean values $\boldsymbol{\mu}$ of the
random vector $\textbf{v}=(x_0\quad y_0\quad \Theta_1\quad\Theta_2)^{\mathrm{T}}$ using Eqs.~(\ref{BW}) and (\ref{ThetaMean})--(\ref{W2covar}) and  knowledge about the corresponding beam, aperture parameters, atmospheric structure constant $C_n^2$, and beam divergence parameter $\Xi$.  We note, however, that the analytical results (\ref{BW}), (\ref{W2mean}), and (\ref{W2covar}) were obtained in the asymptotic case of weak-to-moderate turbulence.
The elliptic-beam approximation in the present form does not work for arbitrary channels.
\item Then the numerical integration in Eq.~(\ref{PDTeq}) can be performed within a Monte Carlo method.  For this
purpose one should simulate the $N$ values of the vector $\textbf{v}$ and the angle $\phi$. The  angle $\phi$ is assumed to be uniformly distributed in the interval $[0,\pi/2]$. The simulated values of  $\textbf{v}$ and $\phi$ are substituted into Eq.~(\ref{etaElBeam}) by taking into account that $\phi{=}\varphi{-}\phi_0$. Finally, the obtained  transmittances are multiplied with the extinction factor $\chi_{\mathrm{ext}}$ yielding $N$ values of atmospheric transmittance  $\chi_{\mathrm{ext}}\eta(\textbf{v}_i,\phi_i)$, $i{=}1,...,N$. The corresponding PDT can be visualized  using the simulated values of transmittance via histograms or using the techniques of smooth kernels~\cite{Wand}.
\item In most practical situations the knowledge of mean value of some quantity that is function of transmittance  $\langle f(\eta)\rangle$ is needed.  Such a quantity is  estimated from the simulated values of transmittance as
\begin{align}
 \langle f(\eta)\rangle\approx\frac{1}{N}\sum_{i=1}^N f(\chi_{\mathrm{ext}}\eta(\textbf{v}_i,\phi_i)).
\end{align}
where $\eta(\textbf{v}_i ,\phi_i)$ is obtained from Eq.~(\ref{etaElBeam}).
For example, one can obtain the first two moments of  the atmospheric transmittance  as
\begin{align}
 \langle\eta\rangle\approx\chi_{\mathrm{ext}}\frac{1}{N}\sum_{i=1}^N \eta(\textbf{v}_i,\phi_i),
\end{align}
\begin{align}
 \langle\eta^2\rangle\approx\chi_{\mathrm{ext}}^2\frac{1}{N}\sum_{i=1}^N \eta^2(\textbf{v}_i,\phi_i).
\end{align}
\end{enumerate}

\section{Statistical parameters for the elliptic beam and optical field correlations } \label{app:PhaseApprox}

The  vector  $\mathbf{v}=(x_0\quad y_0\quad \Theta_1\quad\Theta_2)^T$ is a Gaussian random vector. The angle variable $\phi$ is assumed to be uniformly distributed in the interval $[0,\frac{\pi}{2}]$.  The reference frame is chosen such that $\langle x_0\rangle=\langle y_0\rangle=0$ and
\begin{align}\label{BWvariance}
 \langle x_0^2\rangle=\langle y_0^2\rangle=\frac{1}{\chi_\mathrm{ext}^2}\int_{\mathbb{R}^4}\D^2\boldsymbol{\rho}_1\D^2\boldsymbol{\rho}_2\, x_1x_2\Gamma_4(\boldsymbol{\rho}_1,\boldsymbol{\rho}_2;L),
\end{align}
where $\Gamma_4(\boldsymbol{\rho}_1,\boldsymbol{\rho}_2;z){=}\langle u^\ast(\boldsymbol{\rho}_1,z)u(\boldsymbol{\rho}_1,z)u^\ast(\boldsymbol{\rho}_2,z)u(\boldsymbol{\rho}_2,z)\rangle$ is the fourth-order field-correlation function. The means and (co)variances of $\Theta_i$ are expressed via the means and (co)variances of $W_i^2$ by Eqs.~(\ref{ThetaMean}) and (\ref{ThetaVariance}), respectively. Under the assumptions of Gaussianity and isotropy (for details see the Supplemental Material of Ref.~\cite{Vasylyev2016}) the means and (co)variances of $W_i^2$ read 
\begin{align}\label{BeamBroad}
& \langle W_{1/2}^2\rangle=4\left[\frac{1}{\chi_\mathrm{ext}}\int_{\mathbb{R}^2}\D^2\boldsymbol{\rho}\, x^2 \Gamma_2(\boldsymbol{\rho};L)-\langle x_0^2\rangle\right],\\
\label{Wcovar}
& \langle\Delta W_{i}^2 \Delta W_j^2\rangle={-}\frac{8}{\chi_\mathrm{ext}^2}\Biggl\{2\left(\int_{\mathbb{R}^2}\D^2\boldsymbol{\rho}\, x^2 \Gamma_2(\boldsymbol{\rho};L)\right)^2\Biggr.\nonumber\\
 &{-}\int_{\mathbb{R}^4}\D^2\boldsymbol{\rho}_1\D^2\boldsymbol{\rho}_2\bigl[x_1^2 x_2^2(4\delta_{ij}{-}1)-x_1^2y_2^2(4\delta_{ij}{-}3)\bigr]\nonumber\\
 &\qquad{\times}\Gamma_4(\boldsymbol{\rho}_1,\boldsymbol{\rho}_2;L)\Biggr\}-16\left[4\delta_{ij}-1\right]\langle x_0^2\rangle^2,
\end{align}
where we have also used the second-order field-correlation function $\Gamma_2(\boldsymbol{\rho};z){=}\langle u^\ast(\boldsymbol{\rho},z)u(\boldsymbol{\rho},z)\rangle$.

For the calculation of the field-correlation functions $\Gamma_2$ and $\Gamma_4$ in Eqs.~(\ref{BWvariance})-(\ref{Wcovar}) we use the expression (\ref{uExpression}) for the  field envelope $u(\boldsymbol{\rho},z)$.
Substituting Eq.~(\ref{uExpression}) in the second- and fourth-order field correlation functions and performing the statistical averaging  one gets
\begin{align}\label{Gamma2def}
 &\Gamma_{2}(\boldsymbol{\rho};L)=\chi_{\mathrm{ext}}\int_{\mathbb{R}^4}\D^2\boldsymbol{\rho}_1^\prime\D^2\boldsymbol{\rho}_2^\prime\, u_0(\boldsymbol{\rho}^\prime_1)u^\ast_0(\boldsymbol{\rho}_2^\prime)G_0(\boldsymbol{\rho},\boldsymbol{\rho}^\prime_1;L,0)\nonumber\\
 &{\times} G_0^\ast(\boldsymbol{\rho},\boldsymbol{\rho}^\prime_2;L,0)\left\langle\exp\left[i S(\boldsymbol{\rho},\boldsymbol{\rho}_1^\prime;L,0){-}i S(\boldsymbol{\rho},\boldsymbol{\rho}_2^\prime;L,0)\right]\right\rangle,
\end{align}
\begin{align}
\label{Gamma4def}
 &\Gamma_4(\boldsymbol{\rho}_1,\boldsymbol{\rho}_2;L){=}\eta_{\mathrm{ext}}^2\int_{\mathbb{R}^8}\D^2\boldsymbol{\rho}_1^\prime...\D^2\boldsymbol{\rho}_4^\prime\, u_0(\boldsymbol{\rho}^\prime_1)u^\ast_0(\boldsymbol{\rho}_2^\prime)u_0(\boldsymbol{\rho}^\prime_3)\nonumber\\
 &\times u^\ast_0(\boldsymbol{\rho}_4^\prime) G_0(\boldsymbol{\rho}_1,\boldsymbol{\rho}^\prime_1;L,0)G_0^\ast(\boldsymbol{\rho}_1,\boldsymbol{\rho}^\prime_2;L,0)G_0(\boldsymbol{\rho}_2,\boldsymbol{\rho}^\prime_3;L,0)\nonumber\\
 &\times G_0^\ast(\boldsymbol{\rho}_2,\boldsymbol{\rho}^\prime_4;L,0)\left\langle \exp\left[i S(\boldsymbol{\rho}_1,\boldsymbol{\rho}^\prime_1;L,0){-}i S(\boldsymbol{\rho}_1,\boldsymbol{\rho}^\prime_2;L,0)\nonumber\right.\right.\\
 &\qquad\qquad\qquad \left.\left.+i S(\boldsymbol{\rho}_2,\boldsymbol{\rho}^\prime_3;L,0){-}i S(\boldsymbol{\rho}_2,\boldsymbol{\rho}^\prime_4;L,0)\right]\right\rangle,
\end{align}
with
\begin{align}\label{G0}
 G_0(\boldsymbol{\rho},\boldsymbol{\rho}^\prime;z,z^\prime)=\frac{k}{2\pi i(z-z^\prime)}\exp\left[\frac{ik|\boldsymbol{\rho}-\boldsymbol{\rho}^\prime|^2}{2(z-z^\prime)}\right].
\end{align}
Assuming  that  the relative permittivity $\delta\varepsilon$ is a  Gaussian stochastic field, we can rewrite Eqs.~(\ref{Gamma2def})  and
(\ref{Gamma4def}) as
\begin{align}\label{Gamma2def1}
 &\Gamma_{2}(\boldsymbol{\rho};L)=\chi_{\mathrm{ext}}\int_{\mathbb{R}^4}\D^2\boldsymbol{\rho}_1^\prime\D^2\boldsymbol{\rho}_2^\prime\, u_0(\boldsymbol{\rho}^\prime_1)u^\ast_0(\boldsymbol{\rho}_2^\prime)G_0(\boldsymbol{\rho},\boldsymbol{\rho}^\prime_1;L,0)\nonumber\\
 &\quad\times G_0^\ast(\boldsymbol{\rho},\boldsymbol{\rho}^\prime_2;L,0)\exp\Bigl[-\frac{1}{2}\mathcal{D}_{S}(0,\boldsymbol{\rho}_1^\prime{-}\boldsymbol{\rho}_2^\prime)\Bigr],
\end{align}
\begin{align}
\label{Gamma4def1}
 &\Gamma_4(\boldsymbol{\rho}_1,\boldsymbol{\rho}_2;L){=}\chi_{\mathrm{ext}}^2\int_{\mathbb{R}^8}\!\!\D^2\boldsymbol{\rho}_1^\prime...\D^2\boldsymbol{\rho}_4^\prime\, u_0(\boldsymbol{\rho}^\prime_1)u^\ast_0(\boldsymbol{\rho}_2^\prime)u_0(\boldsymbol{\rho}^\prime_3)u^\ast_0(\boldsymbol{\rho}_4^\prime)\nonumber\\
 &\times G_0(\boldsymbol{\rho}_1,\boldsymbol{\rho}^\prime_1;L,0)G_0^\ast(\boldsymbol{\rho}_1,\boldsymbol{\rho}^\prime_2;L,0)G_0(\boldsymbol{\rho}_2,\boldsymbol{\rho}^\prime_3;L,0)\nonumber\\
 &\times  G_0^\ast(\boldsymbol{\rho}_2,\boldsymbol{\rho}^\prime_5;L,0)\exp\Biggl[-\frac{1}{2}\Bigl\{ \mathcal{D}_{S}(0,\boldsymbol{\rho}_1^\prime{-}\boldsymbol{\rho}^\prime_2)+ \mathcal{D}_{S}(0,\boldsymbol{\rho}_3^\prime{-}\boldsymbol{\rho}^\prime_4)\nonumber\\
 &\qquad\qquad-\sum_{\substack{i=1,2 \\ j=3,4}}(-1)^{i+j}\mathcal{D}_{S}(\boldsymbol{\rho}_1{-}\boldsymbol{\rho}_2,\boldsymbol{\rho}_i^\prime{-}\boldsymbol{\rho}^\prime_j)\Bigr\}\Biggr],
\end{align}
where
\begin{align}
\label{StructFunc}
& \mathcal{D}_{S}(\boldsymbol{\rho}_k{-}\boldsymbol{\rho}_l,\boldsymbol{\rho}_k^\prime{-}\boldsymbol{\rho}_l^\prime)\nonumber\\
&\qquad=\left\langle\left[S(\boldsymbol{\rho}_k,\boldsymbol{\rho}_k^\prime;z,z^\prime)-S(\boldsymbol{\rho}_l,\boldsymbol{\rho}_l^\prime;z,z^\prime)\right]^2\right\rangle
\end{align}
is the structure function of the phase fluctuations.

Using Eq.~(\ref{RandPh}) and the Markovian approximation (cf., e.g., Ref~\cite{Fante}),  we obtain for $z^\prime{=}0$, and $z{=}L$,
\begin{align}\label{DsGeneral}
& \mathcal{D}_{S}(\boldsymbol{\rho}_k{-}\boldsymbol{\rho}_l,\boldsymbol{\rho}_k^\prime{-}\boldsymbol{\rho}_l^\prime)\nonumber\\
&=\frac{k^2L^2}{4}\int_0^1\D\xi\Bigl\langle\Bigl\{\delta \varepsilon\left(\boldsymbol{\rho}_k\xi{+}\boldsymbol{\rho}^\prime_k[1{-}\xi],\xi\right)\nonumber\\
&\qquad\qquad\qquad-\delta \varepsilon\left(\boldsymbol{\rho}_l\xi{+}\boldsymbol{\rho}_l^\prime[1{-}\xi],\xi\right)\Bigr\}^2\Bigr\rangle,
\end{align}
i.e., we assume that turbulent inhomogeneities as well as random scatterers represented by  the relative permittivity $\delta \varepsilon$   are $\delta$ correlated in the $z$ direction. In this case the structure function (\ref{DsGeneral}) can be written in terms of the permittivity fluctuation spectrum  $\Phi_\varepsilon(\boldsymbol{\kappa})$ as
\begin{align}
 &\mathcal{D}_{S}(\boldsymbol{\rho},\boldsymbol{\rho}^\prime)=\frac{\pi}{2}k^2L\int_0^1\D\xi\int_{\mathbb{R}^2}\D^2\boldsymbol{\kappa}\,\Phi_\varepsilon(\boldsymbol{\kappa})\nonumber\\
 &\times\Bigl(1-\exp\bigl\{i\boldsymbol{\kappa}{\cdot}\left[\boldsymbol{\rho}\xi{+}\boldsymbol{\rho}^\prime(1{-}\xi)\right]\bigr\}\Bigr),
\end{align}
where, due to the Markovian approximation, the spectrum depends on the reduced vector $\boldsymbol{\kappa}$, which is related to the vector $\mathbf{K}$ in Eq.~(\ref{PhiSpectrum}) as  $\mathbf{K}=(\kappa_x\quad\kappa_y\quad0)^T$.
Moreover, taking into account that the spectrum splits into two parts  [cf.~Eq.~(\ref{PhiSpectrum})], we can write
\begin{align}
 \mathcal{D}_{S}=\mathcal{D}_{S}^{\mathrm{turb}}+\mathcal{D}_{S}^{\mathrm{scat}},
\end{align}
i.e. the phase structure function also splits into turbulent- and random scattering-induced contributions.

Based on the Kolmogorov turbulence spectrum (\ref{PhiTurb}) and the proposed Gaussian spectrum (\ref{PhiScat}) for random scatterers, we obtain for the corresponding structure functions
\begin{align}\label{StrFunct1}
& \mathcal{D}_{S}^{\rm turb}(\boldsymbol{\rho},\boldsymbol{\rho}^\prime)=2.4 \sigma_R^2k^{\frac{5}{6}}L^{-\frac{5}{6}}\int_0^1\D\xi\left|\boldsymbol{\rho}\xi+\boldsymbol{\rho}^\prime(1-\xi)\right|^{\frac{5}{3}},\\
% \end{align}
% \begin{align}
 &\mathcal{D}_S^{\mathrm{scat}}(\boldsymbol{\rho},\boldsymbol{\rho}^\prime)=2\sigma_{S,\mathrm{scat}}^{2} \int_0^1\D\xi\label{PhaseScrModel}\\%
 &\times\Bigl\{1-\exp\left[-\Bigl|(\boldsymbol{\rho}_1{-}\boldsymbol{\rho}_2)\xi{+}(\boldsymbol{\rho}_1^\prime{-}\boldsymbol{\rho}_2^\prime)(1{-}\xi)\Bigr|^2/4\zeta_0^2\right]\Bigr\}.\nonumber
\end{align}
Here,  the Rytov variance $\sigma_R^2$ is given by Eq.~(\ref{Rytov}),  $\zeta_0$  is the transversal correlation length for random scatterers, and $\sigma_{S,\mathrm{scat}}^{2}$ is the corresponding phase variance given by Eq.~(\ref{PhVar}).
For  weather conditions with high visibility (haze  and thin fog) the correlation length $\zeta_0$ is large. In this case the phase structure function reads
\begin{align}\label{StrFunct2}
 \mathcal{D}_{S}^{\mathrm{scat}}(\boldsymbol{\rho},\boldsymbol{\rho}^\prime){=}\frac{\sigma_{S,\mathrm{scat}}^{2}}{2\zeta_0^2}\int_0^1\D\xi\left|\boldsymbol{\rho}\xi{+}\boldsymbol{\rho}^\prime(1{-}\xi)\right|^2,
\end{align}
and it is a quadratic function of its arguments.

We substitute Eqs.~(\ref{StrFunct1}) and (\ref{StrFunct2}) into Eqs.~(\ref{Gamma2def}) and (\ref{Gamma4def}) and perform the corresponding integration. This results in the  expressions for the field correlation functions
\begin{align}\label{Gamma2def2}
 \Gamma_2(\boldsymbol{\rho})&=\chi_\mathrm{ext}\frac{\Omega^2}{\pi^2W_0^4}\int_{\mathbb{R}^2}\D^2\boldsymbol{\rho}^\prime e^{-\frac{\gamma^2}{2W_0^2}|\boldsymbol{\rho}^\prime|^2-2 i\frac{\Omega}{W_0}\boldsymbol{\rho}\cdot\boldsymbol{\rho}^\prime}\nonumber\\
 &\times\exp\Bigl[-\frac{1}{2}\mathcal{D}_{S}^{\mathrm{turb}}(0,\boldsymbol{\rho}^\prime)\Bigr],
\end{align}
\begin{align}\label{Gamma4def2}
 &\Gamma_4(\boldsymbol{\rho}_1,\boldsymbol{\rho}_2)=\chi_\mathrm{ext}^2\frac{4\Omega^4}{\pi^5W_0^{10}}\int_{\mathbb{R}^6}\D^2\boldsymbol{\rho}^\prime_1
\D^2\boldsymbol{\rho}^\prime_2\D^2\boldsymbol{\rho}^\prime_3\nonumber\\
&\times e^{-\frac{1}{W_0^2}\bigl(|\boldsymbol{\rho}_1^\prime|^2+|\boldsymbol{\rho}_2^\prime|^2+\gamma^2|\boldsymbol{\rho}_3^\prime|^2\bigr)}e^{2i\frac{\Omega}{W_0^2}\boldsymbol{\rho}_1^\prime\cdot\boldsymbol{\rho}_2^\prime}\nonumber\\
&\qquad\times e^{-2i\frac{\Omega}{W_0^2}\bigl[(\boldsymbol{\rho}_1-\boldsymbol{\rho}_2)\cdot\boldsymbol{\rho}_2^\prime+(\boldsymbol{\rho}_1{+}\boldsymbol{\rho}_2)\cdot\boldsymbol{\rho}_3^\prime\bigr]}\\
&\times\exp\Bigl[-\frac{1}{2}\sum_{j=1,2}\Bigl\{\mathcal{D}_{S}^{\mathrm{turb}}(\boldsymbol{\rho}_1{-}\boldsymbol{\rho}_2,\boldsymbol{\rho}_1^\prime{+}(-1)^j\boldsymbol{\rho}_3^\prime)\nonumber\\
&+\mathcal{D}_{S}^{\mathrm{turb}}(0,\boldsymbol{\rho}_2^\prime{+}(-1)^j\boldsymbol{\rho}_3^\prime){-}\mathcal{D}_{S}^{\mathrm{turb}}(\boldsymbol{\rho}_1{-}\boldsymbol{\rho}_2,\boldsymbol{\rho}_1^\prime{+}(-1)^j\boldsymbol{\rho}_2^\prime)\Bigr\}\Bigr]\nonumber.
\end{align}
Here
\begin{align}\label{gamma}
 \gamma^2=1+\Omega^2\left(1-\frac{L}{F}\right)^2+\frac{2}{3}\sigma_{S,\mathrm{scat}}^2\frac{W_0^2}{4\zeta_0^2}
\end{align}
is the generalized beam diffraction parameter that includes the contribution from random scattering and $\Omega$ is the Fresnel number of the transmitter aperture.

\section{Beam wandering and beam shape distortion in the presence of random scatterers and turbulence}\label{app:EllModelParam}

In this appendix we derive the statistical characteristics of the elliptic beam taking into account the presence of random scatterers.  The derivations follow closely the calculations given in the Supplemental Material of Ref.~\cite{Vasylyev2016}.

\subsection{Beam wandering}

The main contribution to   beam wandering comes from large turbulent eddies located close to the beam transmitter~\cite{Andrews}. This allows us
to replace the integral along the propagation path in Eq.~(\ref{DsGeneral}) with its value at the transmitter aperture plane \cite{Kon, Klyatskin, MironovBook}, i.e.,
\begin{align}
\int_{z^\prime}^z\D\xi f(\xi)\approx(z-z^\prime) f(z^\prime).
\end{align}
The phase structure function  then reduces to
\begin{align}
 \mathcal{D}_S(\boldsymbol{\rho},\boldsymbol{\rho}^\prime)= \mathcal{D}_{S}^{\mathrm{turb}}(\boldsymbol{\rho},\boldsymbol{\rho}^\prime)=2.4\sigma_R^2 k^{\frac{5}{6}}L^{-\frac{5}{6}}|\boldsymbol{\rho}^\prime|^{\frac{5}{3}}.
\end{align}
Here we have set $\mathcal{D}_{S}^{\mathrm{scat}}(\boldsymbol{\rho},\boldsymbol{\rho}^\prime){=}0$, which is justified if the characteristic sizes of random scatterers are less than the characteristic sizes of eddies contributing to beam wandering.

Substituting Eq.~(\ref{Gamma4def2}) into (\ref{BWvariance}), we obtain for the beam wandering variance the expression
\begin{align}
 &\langle x_0^2\rangle{=}\frac{4\Omega^4}{\pi^5 W_0^{10}}\int_{\mathbb{R}^8}\D^2\widetilde{\mathbf{R}}\,\D^2\widetilde{\boldsymbol{\rho}}\,\D^2\boldsymbol{\rho}_1^\prime\, \D^2\boldsymbol{\rho}_2^\prime\, \D^2\boldsymbol{\rho}_3^\prime\Bigl(\widetilde{R}_x^2{-}\frac{\widetilde{\rho}_x^2}{4}\Bigr)\nonumber\\
 &{\times} e^{-\frac{1}{W_0^2}\Bigl(|\boldsymbol{\rho}^\prime_1|^2{+}|\boldsymbol{\rho}_2^\prime|^2+g^2|\boldsymbol{\rho}_3^\prime|^2\Bigr)}  e^{-\frac{2i\Omega}{W_0^2}\bigl[\widetilde{\boldsymbol{\rho}}\cdot\boldsymbol{\rho}_2^\prime-\boldsymbol{\rho}_1^\prime\cdot\boldsymbol{\rho}_2^\prime+2\widetilde{\mathbf{R}}\cdot\boldsymbol{\rho}_3^\prime\bigr]}\nonumber\\
 &{\times}\exp\Bigl[-1.2\sigma_R^2k^{\frac{5}{6}}L^{-\frac{5}{6}}\sum_{j=1,2}\Bigl\{|\boldsymbol{\rho}_1^\prime{+}(-1)^j\boldsymbol{\rho}_3^\prime|^{\frac{5}{3}}\Bigr.\Bigr.\\
&\qquad\qquad\Bigl.\Bigl.+|\boldsymbol{\rho}_2^\prime{+}(-1)^j\boldsymbol{\rho}_3^\prime|^{\frac{5}{3}}-|\boldsymbol{\rho}_1^\prime{+}(-1)^j\boldsymbol{\rho}_2^\prime|^{\frac{5}{3}}\Bigr\}\Bigr], \nonumber
\end{align}
where $g^2{=}1{+}\Omega^2[1{-}L/F]^2$ and we have used the variables $\widetilde{\boldsymbol{\rho}}{=}\boldsymbol{\rho}_1{-}\boldsymbol{\rho}_2$ and $\widetilde{\mathbf{R}}{=}(\boldsymbol{\rho}_1{+}\boldsymbol{\rho}_2)/2$.
The integration over the variables $\mathbf{R}$ and $\boldsymbol{\rho}_3^\prime$  can be performed using the properties of the Dirac $\delta$ function, for example, using the relation
\begin{align}
 &\int_{\mathbb{R}^4}\D^2\widetilde{\mathbf{R}}\, \D^2\boldsymbol{\rho}_3^\prime\,\widetilde{\mathbf{R}}^2 e^{-4i\frac{\Omega}{W_0^2}\widetilde{\mathbf{R}}\cdot\boldsymbol{\rho}_3^\prime}f(\boldsymbol{\rho}_3^\prime)\nonumber\\
 &\qquad\qquad=-\frac{(2\pi)^2W_0^8}{(4\Omega)^4}\Delta_{\boldsymbol{\rho}_3^\prime}^2f(\boldsymbol{\rho}_3^\prime)\Bigl.\Bigr|_{\boldsymbol{\rho}_3^\prime=0},
\end{align}
where $\Delta_{\boldsymbol{\rho}_3^\prime}^2$ is the transverse Laplace operator and $f(\boldsymbol{\rho})$ is an arbitrary function. In the limit of weak optical turbulence  ($\sigma_R^2\approx1$)
the integral can be evaluated as
\begin{align}
& \langle x_0^2\rangle{=}\frac{2.4\Omega^2\sigma_R^2k^{\frac{5}{6}}L^{-\frac{5}{6}}}{(2\pi)^3W_0^6}\int_{\mathbb{R}^6}\D^2\widetilde{\boldsymbol{\rho}}\,\D^2\boldsymbol{\rho}_1^\prime\D^2\boldsymbol{\rho}_2^\prime\Biggl(\frac{g^2W_0^2}{2\Omega^2}{-}\widetilde{\rho}_x^2\Biggr)\nonumber\\
 &\qquad\times e^{-\frac{1}{W_0^2}(|\boldsymbol{\rho}_1^\prime|^2+|\boldsymbol{\rho}_2^\prime|^2)}e^{2i\frac{\Omega}{W_0^2}\bigl[\{1-\frac{L}{F}\}\boldsymbol{\rho}_1^\prime\cdot\boldsymbol{\rho}_2^\prime-\widetilde{\boldsymbol{\rho}}\cdot\boldsymbol{\rho}_2^\prime\bigr]}\nonumber\\
 &\qquad\times\Bigl(\sum_{j=1,2}|\boldsymbol{\rho}_1^\prime+(-1)^j\boldsymbol{\rho}_2^\prime|^{\frac{5}{3}}-2|\boldsymbol{\rho}_1^\prime|^{\frac{5}{3}}-2|\boldsymbol{\rho}_2^\prime|^{\frac{5}{3}}\Bigr).
 \end{align}
Performing the multiple integration for a focused beam, $L{=}F$, we obtain Eq.~(\ref{BW}).

\subsection{Beam-shape distortion}

Along the whole propagation path the eddies whose sizes are smaller than comparable to the beam diameter contribute to random beam broadening and beam-shape distortion. Here we show that this additional
broadening and beam-shape distortion arise due to the presence of random scatterers.

The moment $\langle W_{1/2}^2\rangle$ defined by Eq.~(\ref{BeamBroad}) contains the following integral:
\begin{align}\label{IntGamma2}
 &\int_{\mathbb{R}^2}\D^2\boldsymbol{\rho}\, x^2\Gamma_2(\boldsymbol{\rho})=\frac{W_0^2\chi_\mathrm{ext}}{\pi^2\Omega^4}\int_{\mathbb{R}^4}\D^2\boldsymbol{\rho}\D^2\boldsymbol{\rho}^\prime\, x^2 e^{-\frac{\gamma^2}{2\Omega^2}|\boldsymbol{\rho}^\prime|^2}\nonumber\\
 &\times \exp\Bigl[-\frac{2i}{\Omega}\boldsymbol{\rho}\cdot\boldsymbol{\rho}^\prime{-}2.14\sigma_R^2\Omega^{\frac{5}{6}}\int_0^1\D\xi(1{-}\xi)^{\frac{5}{3}}\Bigl(\frac{|\boldsymbol{\rho}^\prime|}{\Omega}\Bigr)^{\frac{5}{3}}\Bigr].
\end{align}
 Here the expression (\ref{Gamma2def2}) was used. The integration in Eq.~(\ref{IntGamma2}) can be performed using the approximation $\Bigl(|\boldsymbol{\rho}^\prime/\Omega|\Bigr)^{\frac{5}{3}}\approx\Bigl(|\boldsymbol{\rho}^\prime/\Omega|\Bigr)^2$ (cf.~Ref.~\cite{Andrews}). The resulting expression for the first moment
of $W_{1/2}^2$ in the case of a focused beam results in Eq.~(\ref{W2mean}).

The (co)variances of $W_{1/2}^2$ defined in Eq.~(\ref{Wcovar}) contain the following integrals (cf. Supplemental Material of Ref.~\cite{Vasylyev2016}):
\begin{align}
 &\int_{\mathbb{R}^4}\D^2\boldsymbol{\rho}_1\D^2\boldsymbol{\rho}_2 x_1^2 x_2^2\Gamma_4(\boldsymbol{\rho}_1,\boldsymbol{\rho}_2)=\frac{\Omega^2\chi_\mathrm{ext}}{2(2\pi)^3W_0^6}\int_{\mathbb{R}^4}\D^2
 \boldsymbol{\rho}\D^2\boldsymbol{\rho}_1^\prime\D^2\boldsymbol{\rho}^\prime_2\nonumber\\
 &\times\Bigl(\frac{3 \gamma^4 W_0^4}{4\Omega^4}-\frac{\gamma^2 W_0^2}{\Omega^2}x^2+x^4\Bigr)e^{-\frac{1}{W_0^2}\bigl(|\boldsymbol{\rho}^\prime_1|^2+|\boldsymbol{\rho}_2^\prime|\bigr)}\nonumber\\
 &\times \exp\Bigl[2i\frac{\Omega}{W_0^2}\bigl(1-\frac{L}{F}\bigr)\boldsymbol{\rho}_1^\prime\cdot\boldsymbol{\rho}_2^\prime-2i\frac{\Omega}{W_0^2}\boldsymbol{\rho}\cdot\boldsymbol{\rho}^\prime_2\Bigr]\\
 &\times\exp\Bigl[-1.2\sigma_R^2k^{\frac{5}{6}}L^{-\frac{5}{6}}\int_0^1\D\xi\Bigl(2|\boldsymbol{\rho}\xi+\boldsymbol{\rho}_1^\prime(1-\xi)|^{\frac{5}{3}}\Bigr.\Bigr.\nonumber\\
 &\Bigl.\Bigl.+2(1-\xi)^{\frac{5}{3}}|\boldsymbol{\rho}_2^\prime|^{\frac{5}{3}}-\sum_{j=1,2}|\boldsymbol{\rho}\xi+[\boldsymbol{\rho}_1^\prime+(-1)^j\boldsymbol{\rho}_2^\prime](1-\xi)|^{\frac{5}{3}}\Bigr)\Bigr],\nonumber
\end{align}
\begin{align}
 &\int_{\mathbb{R}^4}\D^2\boldsymbol{\rho}_1\D^2\boldsymbol{\rho}_2 x_1^2 y_2^2\Gamma_4(\boldsymbol{\rho}_1,\boldsymbol{\rho}_2)=\frac{\Omega^2\chi_\mathrm{ext}^2}{2(2\pi)^3W_0^6}\int_{\mathbb{R}^4}
 \D^2\boldsymbol{\rho}\D^2\boldsymbol{\rho}_1^\prime\D^2\boldsymbol{\rho}^\prime_2\nonumber\\
 &\times\Bigl(\frac{ \gamma^4 W_0^4}{4\Omega^4}+\frac{\gamma^2 W_0^2}{\Omega^2}x^2+x^2y^2\Bigr)e^{-\frac{1}{W_0^2}\bigl(|\boldsymbol{\rho}^\prime_1|^2+|\boldsymbol{\rho}_2^\prime|\bigr)}\nonumber\\
 &\times \exp\Bigl[2i\frac{\Omega}{W_0^2}\bigl(1-\frac{L}{F}\bigr)\boldsymbol{\rho}_1^\prime\cdot\boldsymbol{\rho}_2^\prime-2i\frac{\Omega}{W_0^2}\boldsymbol{\rho}\cdot\boldsymbol{\rho}^\prime_2\Bigr]\\
 &\times\exp\Bigl[-1.2\sigma_R^2k^{\frac{5}{6}}L^{-\frac{5}{6}}\int_0^1\D\xi\Bigl(2|\boldsymbol{\rho}\xi+\boldsymbol{\rho}_1^\prime(1-\xi)|^{\frac{5}{3}}\Bigr.\Bigr.\nonumber\\
 &\Bigl.\Bigl.+2(1-\xi)^{\frac{5}{3}}|\boldsymbol{\rho}_2^\prime|^{\frac{5}{3}}-\sum_{j=1,2}|\boldsymbol{\rho}\xi+[\boldsymbol{\rho}_1^\prime+(-1)^j\boldsymbol{\rho}_2^\prime](1-\xi)|^{\frac{5}{3}}\Bigr)\Bigr].\nonumber
\end{align}
We evaluate the multiple integrals by expanding the last exponents into a series with respect to $\sigma_R^2$ up to the first order. For a focused beam ($L{=}F$) we obtain
\begin{align}\label{xxGamma4}
 \frac{1}{\chi_\mathrm{ext}^2}\int_{\mathbb{R}^4}\D^2&\boldsymbol{\rho}_1\D^2\boldsymbol{\rho}_2  x_1^2 x_2^2\Gamma_4(\boldsymbol{\rho}_1,\boldsymbol{\rho}_2){=}\frac{W_0^4}{16\Omega^4}\bigl[1{+}\frac{2}{3}\sigma_{S,\mathrm{scat}}^{2}\frac{W_0^2}{4\xi_0^2}\bigr]^2\nonumber\\
 &+0.58 W_0^4\bigl[1{+}\frac{2}{3}\sigma_{S,\mathrm{scat}}^{2}\frac{W_0^2}{4\zeta_0^2}\bigr]\sigma_R^2\Omega^{-\frac{19}{6}},
\end{align}
\begin{align}\label{xyGamma4}
  \frac{1}{\chi_\mathrm{ext}^2}\int_{\mathbb{R}^4}&\D^2\boldsymbol{\rho}_1\D^2\boldsymbol{\rho}_2  x_1^2 y_2^2\Gamma_4(\boldsymbol{\rho}_1,\boldsymbol{\rho}_2){=}\frac{W_0^4}{16\Omega^4}\bigl[1{+}\frac{2}{3}\sigma_{S,\mathrm{scat}}^{2}\frac{W_0^2}{4\xi_0^2}\bigr]^2\nonumber\\
 &+0.51 W_0^4\bigl[1{+}\frac{2}{3}\sigma_{S,\mathrm{scat}}^{2}\frac{W_0^2}{4\zeta_0^2}\bigr]\sigma_R^2\Omega^{-\frac{19}{6}}.
\end{align}
Substituting Eqs.~(\ref{BW}), (\ref{W2mean}), (\ref{xxGamma4}), and (\ref{xyGamma4}) into Eq.~(\ref{Wcovar}),
 we obtain Eq.~(\ref{W2covar}).

\end{document}